\documentclass[journal]{IEEEtran}
\usepackage{cite}
\usepackage{amsmath,amssymb,amsfonts}
\usepackage{cuted}
\usepackage{graphicx}
\usepackage{textcomp,nicefrac}
\usepackage{wrapfig}
\usepackage{soul}
\usepackage{listings}
\usepackage{color}
\usepackage{floatrow}
\usepackage{xcolor}
\usepackage{hyperref}
\usepackage{tabularx}
\usepackage{graphicx}
\usepackage{comment}
\usepackage{wasysym}
\usepackage{cancel}
\usepackage{threeparttable}
\usepackage{enumitem}

\usepackage{scalerel}
\usepackage{tikz}
\usetikzlibrary{svg.path}
\definecolor{orcidlogocol}{HTML}{A6CE39}
\tikzset{
  orcidlogo/.pic={
    \fill[orcidlogocol] svg{M256,128c0,70.7-57.3,128-128,128C57.3,256,0,198.7,0,128C0,57.3,57.3,0,128,0C198.7,0,256,57.3,256,128z};
    \fill[white] svg{M86.3,186.2H70.9V79.1h15.4v48.4V186.2z}
                 svg{M108.9,79.1h41.6c39.6,0,57,28.3,57,53.6c0,27.5-21.5,53.6-56.8,53.6h-41.8V79.1z M124.3,172.4h24.5c34.9,0,42.9-26.5,42.9-39.7c0-21.5-13.7-39.7-43.7-39.7h-23.7V172.4z}
                 svg{M88.7,56.8c0,5.5-4.5,10.1-10.1,10.1c-5.6,0-10.1-4.6-10.1-10.1c0-5.6,4.5-10.1,10.1-10.1C84.2,46.7,88.7,51.3,88.7,56.8z};
  }
}

\newcommand\orcidicon[1]{\href{https://orcid.org/#1}{\mbox{\scalerel*{
\begin{tikzpicture}[yscale=-1,transform shape]
\pic{orcidlogo};
\end{tikzpicture}
}{|}}}}

\def\BibTeX{{\rm B\kern-.05em{\sc i\kern-.025em b}\kern-.08em
T\kern-.1667em\lower.7ex\hbox{E}\kern-.125emX}}

\begin{document}

\bibliographystyle{ieeetr} 
\title{Analytic timing calculations and timing limits \\ with prompt photons, high-aspect-ratio crystals,\\ and complex TOF-kernels in TOF-PET}

\author{Nicolaus Kratochwil, %\orcidicon{0000-0001-5297-1878} 
\IEEEmembership{Member, IEEE}, Emilie Roncali, %\orcidicon{0000-0002-2439-1064}
\IEEEmembership{Senior member,~IEEE},  Gerard Ari\~no-Estrada, 
%\orcidicon{0000-0002-6411-191X}
\IEEEmembership{Member, IEEE} \thanks{
%
%Manuscript received 11 September 2024; revised 02 December 2024 and 29 January 2025; accepted 03 February 2025. Date of publication xxxx; date of current version xxxx. 
This work was supported by the National Institute of Health grant R01EB034062 (PI: G. Ari\~no-Estrada).
This work did not involve human subjects or animals in its research.\\ 
N. Kratochwil, E. Roncali, and G. Ari\~no-Estrada are with the Department of Biomedical Engineering, University of California at Davis (UCD), Davis, CA, United States \text{(email:~nkratochwil@ucdavis.edu}).
G. Ari\~no-Estrada is also with the Institut de Física d'Altes Energies - Barcelona Institute of Science and Technology, Bellaterra, Barcelona, Spain. \\
%Color versions of one or more figures in this article are available at xxxx. \\
%Digital Object Identifier xxxx.}
}}

\maketitle
\begin{abstract}
Modeling the timing performance of light-based radiation detectors accurately is essential for optimizing time-of-flight positron emission tomography (TOF-PET).
We present an analytic framework that combines existing models to predict the timing behavior of high-aspect ratio crystals, including contributions from prompt photons such as Cherenkov radiation.
This framework is built on a closed-form solution for optical light transport, convolved with the photodetector response and photon production characteristics. 
Using conditional and joint probability distributions, we compute the first-photon arrival time distribution for hybrid detectors with scintillation and Cherenkov light.
The detection time distribution is then self-convolved to derive the time delay spectra and three timing metrics are used to characterize complex TOF kernels.
Additionally, we perform Cramér-Rao Lower Bound  calculations with and without depth-of-interaction bias to evaluate the theoretical timing limits.
Our analytic predictions align well with Monte Carlo simulations for BGO detectors under varying crystal thicknesses and single photon time resolution considering a digital photodetector.
We show that the TOF shape is significantly affected by prompt photon statistics, crystal thickness, scintillation yield, and photodetector properties resulting in distinct metric-dependent timing performance.
The proposed model enables rapid timing predictions for polished crystals, with the calculation time of a detector configuration in under a second, allowing for comprehensive parametric studies.
This makes it a powerful tool for guiding detector development in fast-timing applications.
\end{abstract}

\begin{IEEEkeywords}
time-of-flight PET, coincidence time resolution, radiation detectors, Cherenkov emission, analytic model, Cramér-Rao Lower Bound (CRLB), depth of interaction (DOI)
\end{IEEEkeywords}
%_________________
%TODO notes:
% -remove DE and SE acronym
% work on the formatting, no idea why but somethimes overleaf does not behave well
%______________
\section{Introduction}
\label{sec:intro}
 \IEEEPARstart{M}{odeling} and predicting the timing performance for light-based radiation detection systems is a powerful tool to guide R\&D efforts in fast timing applications such as time-of-flight positron emission tomography (TOF-PET).
 Due to the complexity of the detection system, no closed-form expression exists that covers the full radiation detection chain consisting of photon emission, light transport and detection, photodetector response, readout electronics, and digitization.
Monte Carlo simulations are common tools~\cite{Levin_1996_IEEE,GENTIT_NIMA_2002,ALLISON_2016_NIMA,Sarrut_2021_PMB} to predict the performance and investigate dependencies, such as the impact of light transport and prompt photon emission on the timing capability of a scintillator~\cite{Razdevsek_2023_TRPMS}.
 While such simulations can accurately predict the photon detection time and statistics, computational complexity and computation time can scale up significantly with the number of simulated particles, or with required small step size~\cite{Trigila_FPhys_2022}.
Work is ongoing to accelerate simulations, for instance with generative artificial intelligence~\cite{Trigila_2023_MLST}. 

Attempts have been made to reduce the complexity and focus on dedicated tasks, so that analytic approximations and theoretical frameworks can be used.
In the pioneering work of Seifert et al~\cite{Seifert_2012_PMB}, the Cramer-Rao formalism is applied to scintillation detectors to calculate the intrinsic limit (Cramér-Rao Lower Bound, CRLB) on the timing resolution.
Later Gundacker et al~\cite{Gundacker_2013_JINST} extended this idea to include light transport (obtained with simulations~\cite{Vinke_2014_PMB}) and to mimic the output signal of analog SiPMs.
In the work of Cates et al~\cite{Cates_2015_PMB} an analytic closed-form expression for light transport with single-ended readout of high-aspect-ratio was presented.
The production, detection and timing impact of prompt photons on top of scintillation photons was studied in Gundacker et al~\cite{Gundacker_2016_PMB} and CRLB calculations for different number of prompt photons, as well as photodetector properties, were evaluated.
Further, Vinogradov~\cite{VINOGRADOV_2018_NIMA} presented analytic approximations of coincidence time resolution for scintillation detectors with leading edge discrimination, which was experimentally validated in~\cite{Gundacker_2020_PMB}.
Typically, only a handful of prompt Cherenkov photons are available in TOF-PET detectors, which creates event to event fluctuations on the associated timing capabilities~\cite{Kratochwil_2020_PMB}.
The importance of considering those fluctuations was recently theoretically demonstrated by Loignon-Houle et al~\cite{Loignon-Houle_2023_TRPMS}.

Non-negligible light transport~\cite{Loignon-Houle_2021_PMB,Kratochwil_2021_PMB}, depth-of-interaction (DOI) bias~\cite{Toussaint_2019_PMB,Efthimiou_2020_EJNMMI}, photodetector noise~\cite{Kratochwil_SensorsActivations_2025}, and in general photodetectors with fast response combined with low jitter readout electronics~\cite{Cates_2018_PMB,Gundacker_2019_PMB,Merzi_2023_JINST,Penna_IEEE_2024,Lee_2024_TRPMS} can lead to non-Gaussian TOF distributions.
Pronounced tails in the time delay histogram due to the detection of scintillation light is especially relevant in the framework of scintillation-Cherenkov detectors~\cite{Kwon_2016_PMB,Brunner_2017_PMB,Kratochwil_TRPMS_2024_TlCl}. 
Certain detector configurations~\cite{Ota_2019_PMB,Nadig_2022_TRPMS} can also exhibit side peaks which must not be ignored.
Nuyts et al~\cite{Nuyts-TMI-2023} showed that for any general (non-Gaussian) TOF distribution an associated signal-to-noise (SNR) gain for TOF-PET image reconstruction can be calculated (within certain assumptions outlined in the original manuscript).
We have simplified this concept in ~\cite{Kratochwil_TRPMS_2025_dual} to convert the SNR gain of complex TOF distributions to the SNR gain of a single Gaussian distribution with a given full width at half maximum (FWHM) (= CTR (SNR)).

Among others, these mentioned efforts have contributed to a better understanding of light-based radiation detectors and associated timing limits. 
However, with increasing interest of the TOF-PET community in using prompt photons, and as a promising pathway toward the 10~ps TOF challenge~\cite{Lecoq_2020_PMB}, a revisit and extension of the aforementioned theoretical work can bring valuable insights for fast timing with prompt photons.
This extension is in particular important in view of non-negligible and non-Gaussian light transport jitter for prompt photons in high-aspect ratio crystals, which is progressing to become a main bottleneck for ultra-fast timing.

The first aim of this manuscript is to introduce a combination of theoretical timing performance estimations which integrates key aspects (scintillation and Cherenkov light, high-aspect-ratio crystals, DOI-bias, prompt photon fluctuations, photodetector properties, non-Gaussian TOF shape, CRLB calculations) mentioned before.
A core difference from the aforementioned studies is the use of conditional probability density functions and convolutions of these functions to derive the final TOF-distribution shape of a given detector.
Three different statistical metrics (FWHM, SNR (CTR), standard deviation are used to quantify the shape of the distribution produced by the first detected photon.
In addition, CRLB calculations are performed considering the information of all detected photons.
Secondly, we apply our analytic model to discuss the impact of crystal properties, prompt photons, and photodetector properties on the timing limits and TOF-shape.

The underlying numeric framework for all calculations is written in C++ and built in a modular structure.
This means current limitations and approximations, such as the assumed perfectly polished crystal interface without lateral light loss, can be substituted with a more refined modeling and extended when deemed necessary.
Further, the photodetector is considered to be digital. 
Each photon's timestamp can be captured without readout electronics contribution, photodetector noise, or bandwidth limitation.

\section{Methods}

\subsection{Detector}
We consider a single-ended readout scintillating crystal with thickness $L$ and refractive index $n$, perfectly polished and surrounded by air ($n_A$ = 1.0).
One end is coupled to the photodetector with Cargille Meltmount ($n_M$ = 1.582), while the other end is covered by a specular reflector (ESR) with reflectivity of $R_r =98\%$.
The scintillation emission is material dependent and defined by a single scintillation rise time $\tau_r$ and multiple scintillation decay times $\tau_{d,i}$ with relative abundance $R_i$ ($\sum_iR_i=1$).
The total number of detected photons $M$ is defined as the product of the intrinsic light yield ($LY$), the light transfer efficiency ($LTE$ = ratio of produced and detected photons with a perfect photodetector), and the weighted scintillation detection efficiency ($PDE_{scint}$).
Cherenkov emission is considered as prompt~\cite{Cherenkov_1937} (Dirac-delta function) and isotropic~\cite{Brunner_2014_TNS,Trigila_FPhys_2022}.
The mean number of detected Cherenkov photons is the product of the number of produced Cherenkov photons ($N_{Cher}$), the weighted Cherenkov detection efficiency ($PDE_{Cher}$), and the LTE.
If not stated otherwise, the default material is BGO.
Input parameters for BGO and all tested materials are provided in table~\ref{tab:crystals}.

\subsection{Light propagation and depth-of-interaction}
\label{sec:light-propagation}

Optical photons produced at the interaction position $DOI$ are modeled to have an isotropic angular distribution.
Photons may be forward oriented (toward the photodetector) and within the extraction cone, sidewards oriented (escape to the air, or trapped in the crystal), or backwards oriented to be reflected by the ESR and eventually detected.
We use the analytic expression originally published by Cates et al~\cite{Cates_2015_PMB} (equations 2 \& 3) and adapted in~\cite{Kratochwil_TRPMS_2025_dual} to describe the probability of a photon to travel between a distance of $x$ and $x+dx$.
This is the sum of the forward and backward probability as expressed in equation~\ref{eq:transport}, with $\theta_{CM}$ and $\theta_{CA}$ being the critical angle between the crystal and Meltmount or crystal and air, respectively, while $R_f$ are losses due to Fresnel reflections at the interface between the crystal and Meltmount.

\begin{figure}[h!] 
\begin{center}
    \includegraphics[width=0.49\textwidth]{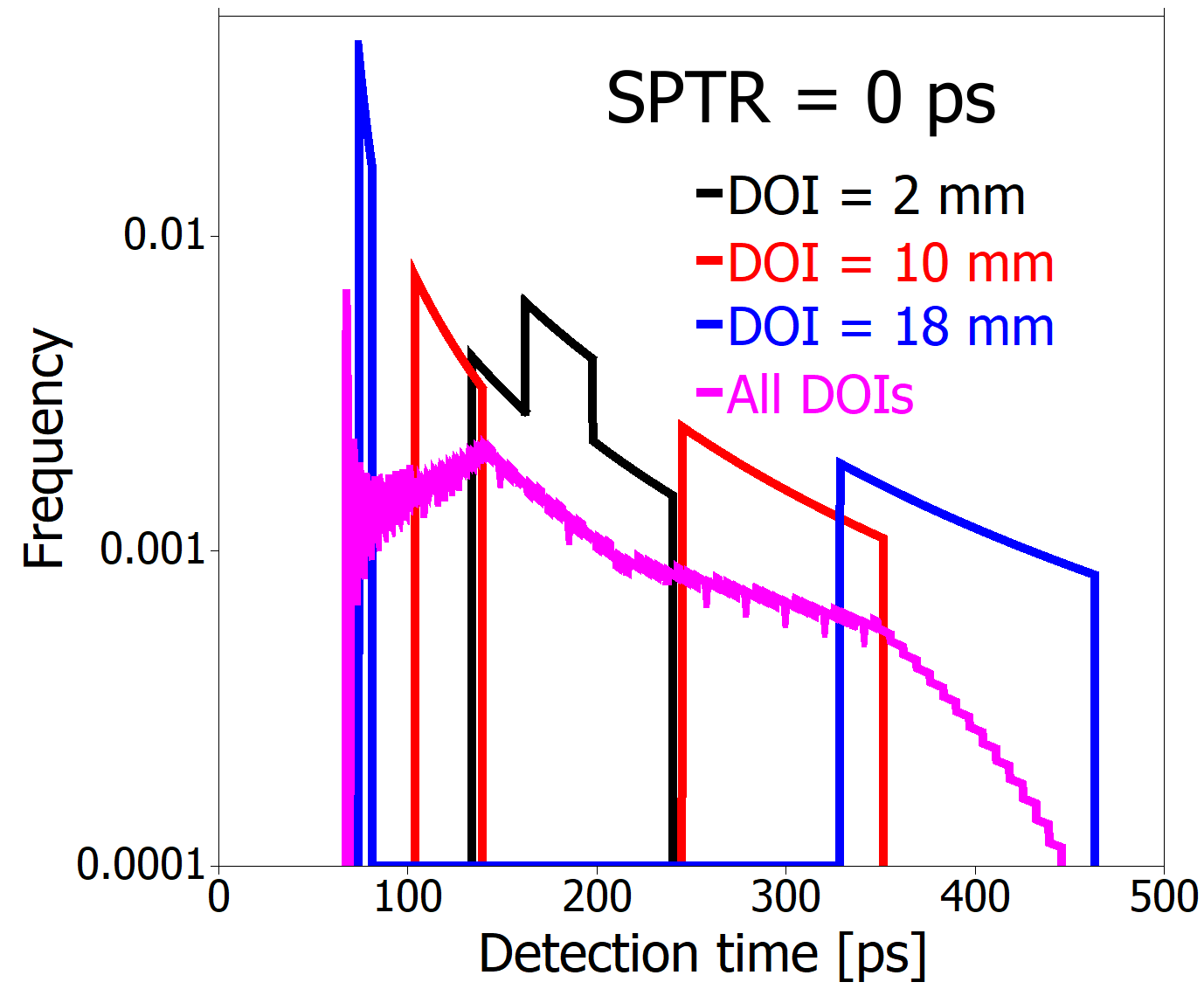}   
    \includegraphics[width=0.49\textwidth]{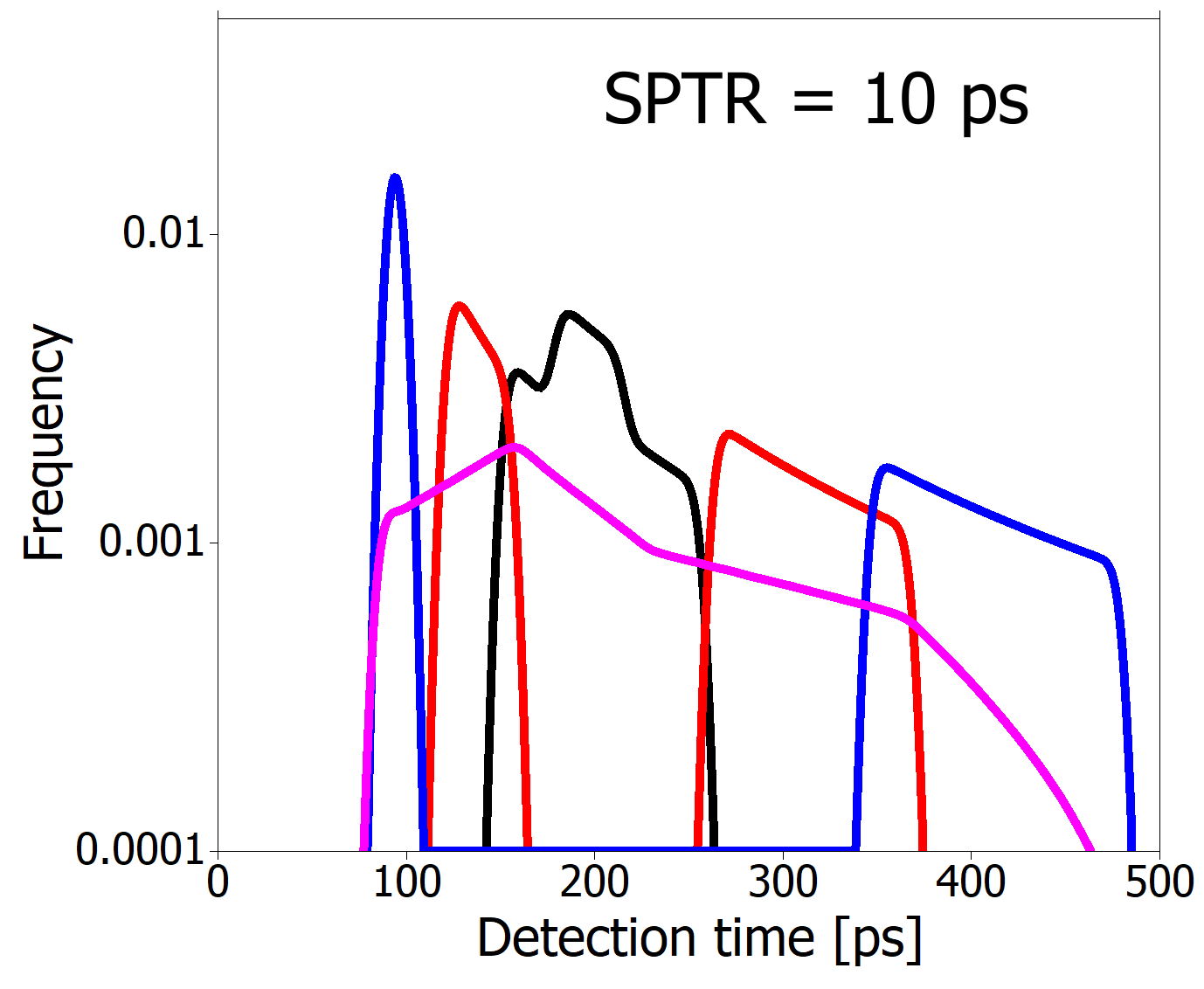} \\
        \includegraphics[width=0.49\textwidth]{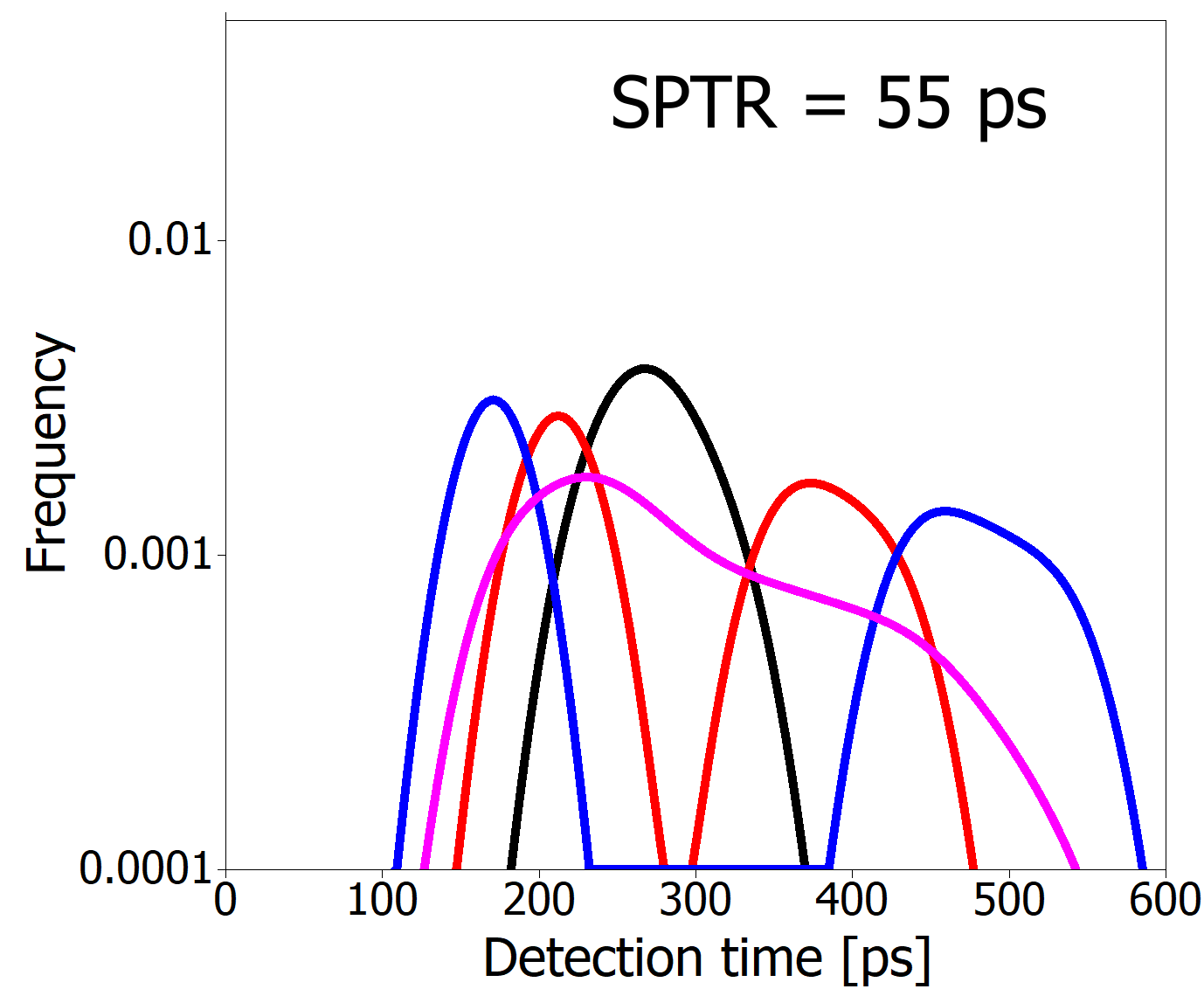}   
    \includegraphics[width=0.49\textwidth]{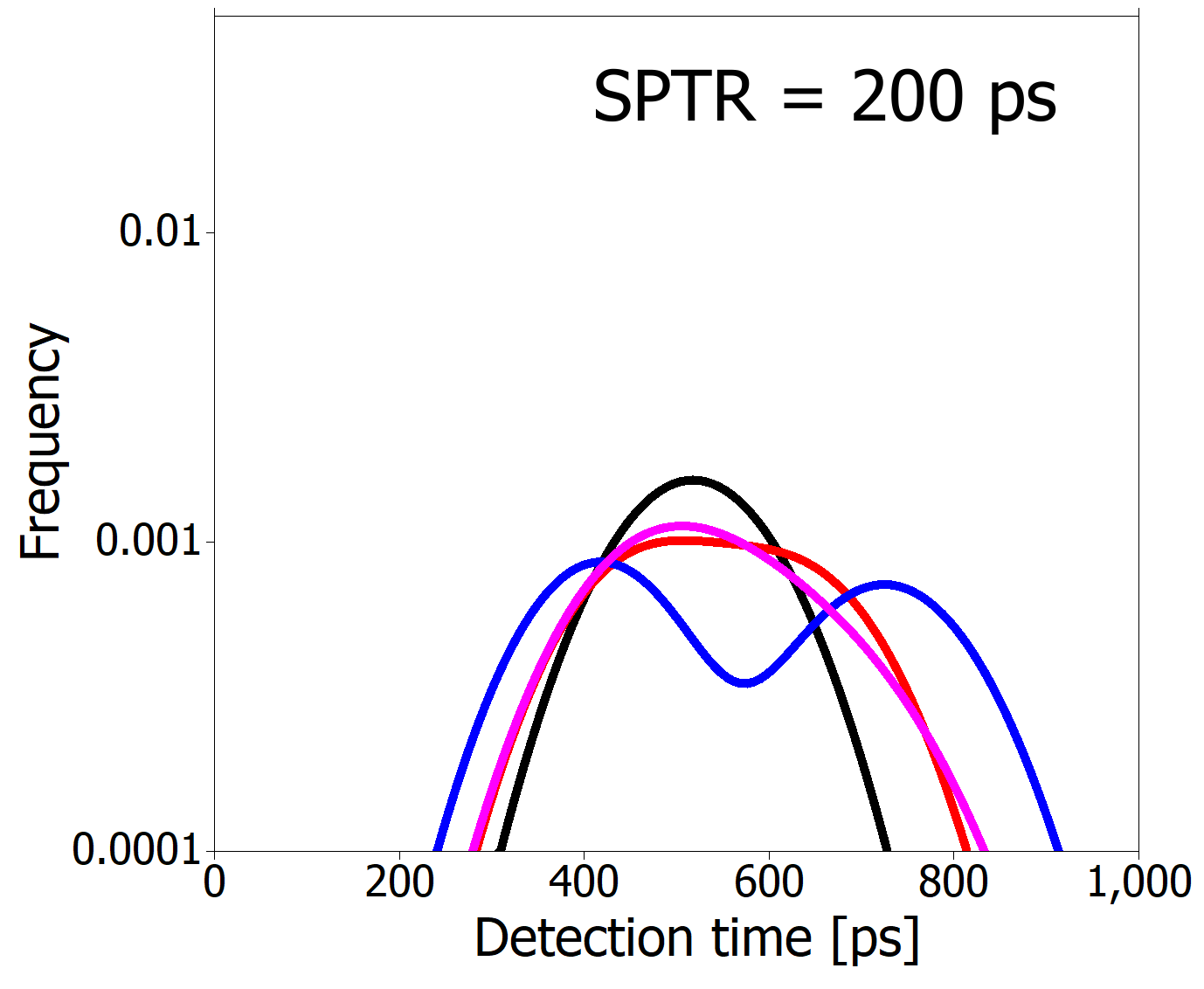} \\
\end{center}
\caption{Photon detection time probability distribution for one prompt photon for different DOIs and SPTR. The scale of the x-axis is increased for higher SPTR values (bottom figures) to cover the full range. The spikes in the detection time for all DOIs without SPTR are due to limited DOI sampling (first DOI = 0.25~mm, last DOI = 19.75~mm, $dDOI$ = 0.5~mm).}
\label{fig:SPTR}
\end{figure}

Once a photon enters Meltmount, it is considered detected.
The integral over all possible distances (hence the fraction of solid angles where light reaches the photodetector) yields the LTE.
Given the perfectly-polished-crystal-approximation without light loss due to bulk absorption or loss upon reflection on the lateral surface (the model is invariant to the crystal pitch), there is little impact of the LTE on the thickness or DOI.
Therefore, we evaluate the LTE once for a 20~mm thick crystal at center DOI and keep it constant for the specific material and all DOIs and thicknesses.
The photon path is normalized to generate a probability distribution vector, whereas the spacing equals $dx$ or, converting into time coordinates, $dt = dx \cdot n / c$.
Additionally, a DOI dependent time shift ($DOI/(c\cdot dt)$) is added to consider the different gamma travel time in the crystal before the interaction.
The top left of figure~\ref{fig:SPTR} shows the detection time for different DOIs and the respective shift due to the gamma travel time and time of the optical photon to reach the photodetector.
\begin{strip}
\vspace{-10mm}
\begin{equation}
\begin{split}
& f(x)_{forward} = 
\begin{cases}
0 & \text{for }  x < (L-DOI)  \text{ and } x > \frac{(L-DOI)}{\text{sin}\left(\pi/2 - \theta_{CM} \right)}\\
\left(\frac{x}{x-dx} -1 \right) \cdot \frac{(L-DOI) \left(1-R_f \right)}{2  x} & \text{for }(L-DOI) \leq x \leq \frac{(L-DOI)}{\text{sin}\left(\pi/2 - \theta_{CM} \right)} 
\end{cases} 
\\
& f(x)_{backward} = 
\begin{cases}
0 & \text{for }  x < (L+DOI)) \text{ and } x > \frac{L+DOI}{\text{sin}\left(\pi/2 - \theta_{CM} \right)} \\
\left(\frac{x}{x-dx} -1 \right) \cdot \frac{(L+DOI) R_r}{2  x} & \text{for }L+DOI \leq x \leq \frac{L+DOI}{\text{cos}\left(\theta_{CA} \right)} \\
\left(\frac{x}{x-dx} -1 \right) \cdot \frac{(L+DOI) (1-R_f)}{2  x} & \text{for } \frac{L+DOI}{\text{cos}\left(\theta_{CA} \right)} < x \leq \frac{L+DOI}{\text{sin}\left(\pi/2 - \theta_{CM} \right)} 
\end{cases} 
\\
& f(x)_{combined} = f(x)_{forward} + f(x)_{backward}
\end{split}
\label{eq:transport}
\end{equation}
\end{strip}
\subsection{Photodetector single photon time resolution}
The time response of the photodetector (single photon time resolution (SPTR) for a SiPM, transit time spread (TTS) for a (micro-channel-plate) photomultiplier tube, (MCP-) PMT) is defined as a Gaussian distribution with a given FWHM.
The centroid is shifted by 4 times the standard deviation to ensure its symmetry in both directions while avoiding negative time values.
This absolute shift has no impact on the calculations.
The light transport is convolved with the photodetector response as illustrated in figure~\ref{fig:SPTR} with different SPTR values.
Small SPTR values preserve the shape of the light transport, while large SPTR values broaden the distribution and lead to a more Gaussian-like shape.
If not stated otherwise, the default SPTR value is 55~ps FWHM to match with the NUV-MT SiPM technology~\cite{Lee_2024_TRPMS}.
\subsection{Generation of scintillation photons}
\label{sec:scintillation}
\begin{figure}[h!] 
\begin{center}
    \includegraphics[width=0.49\textwidth]{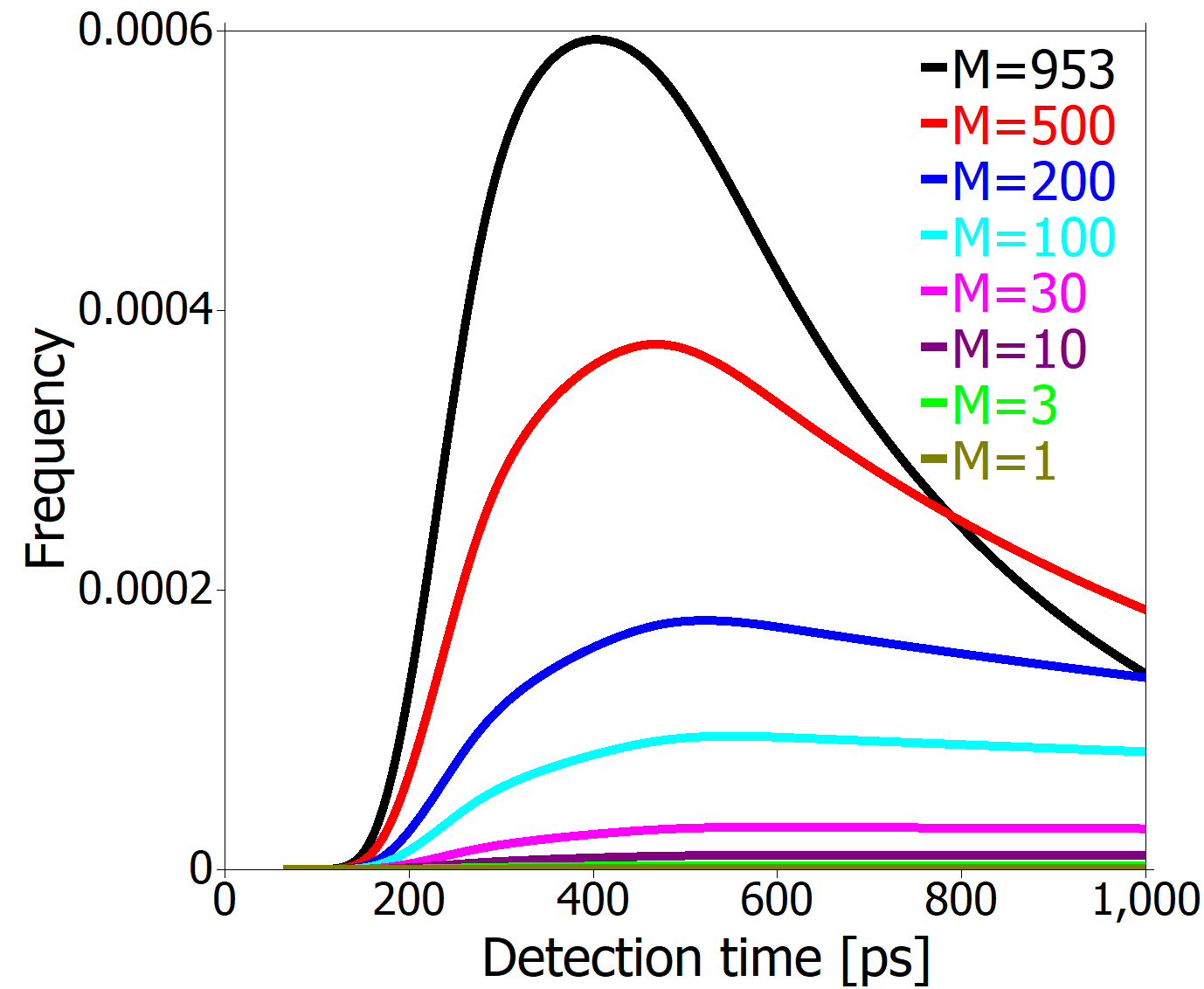}   
    \includegraphics[width=0.49\textwidth]{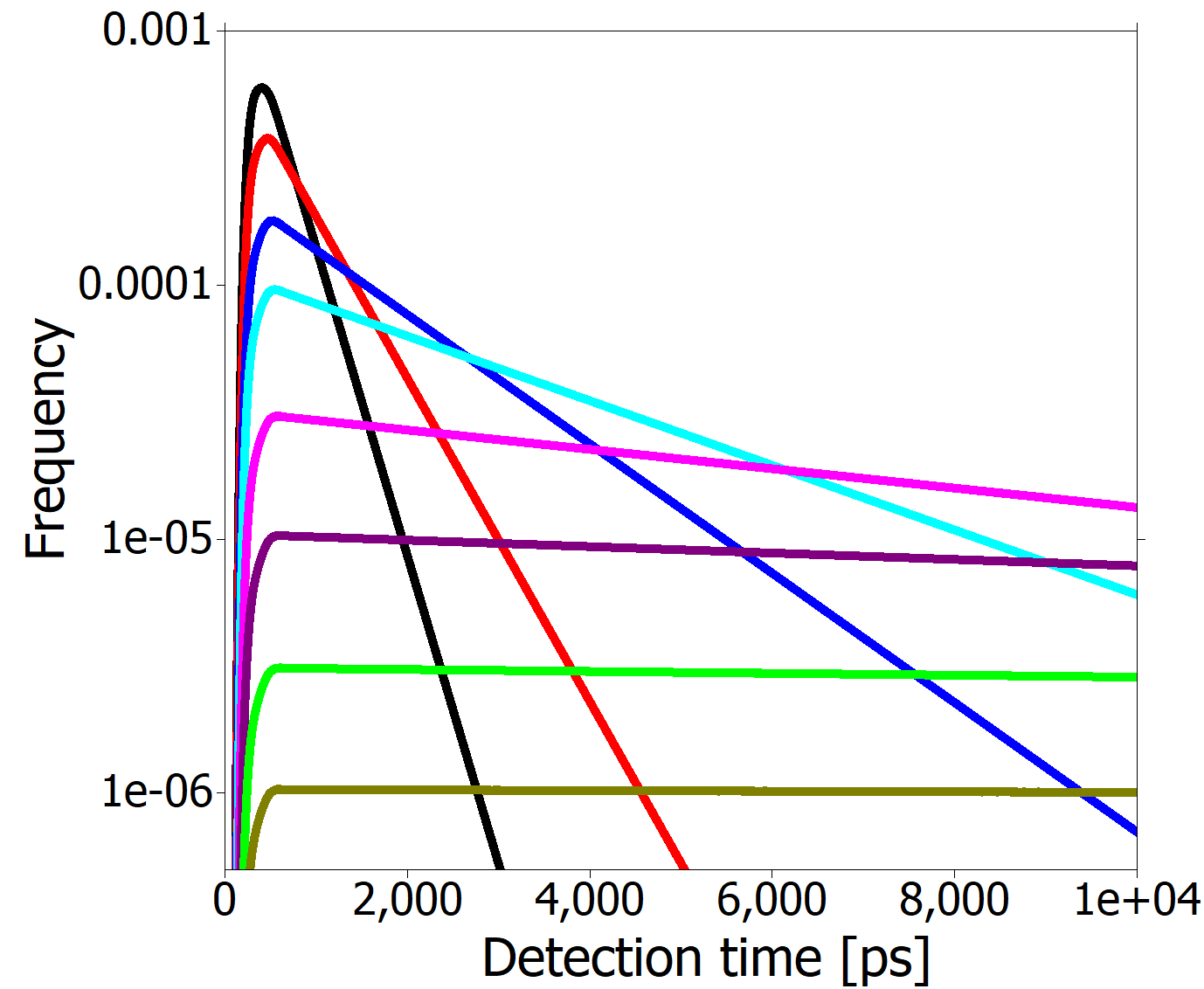} \\
        \includegraphics[width=0.49\textwidth]{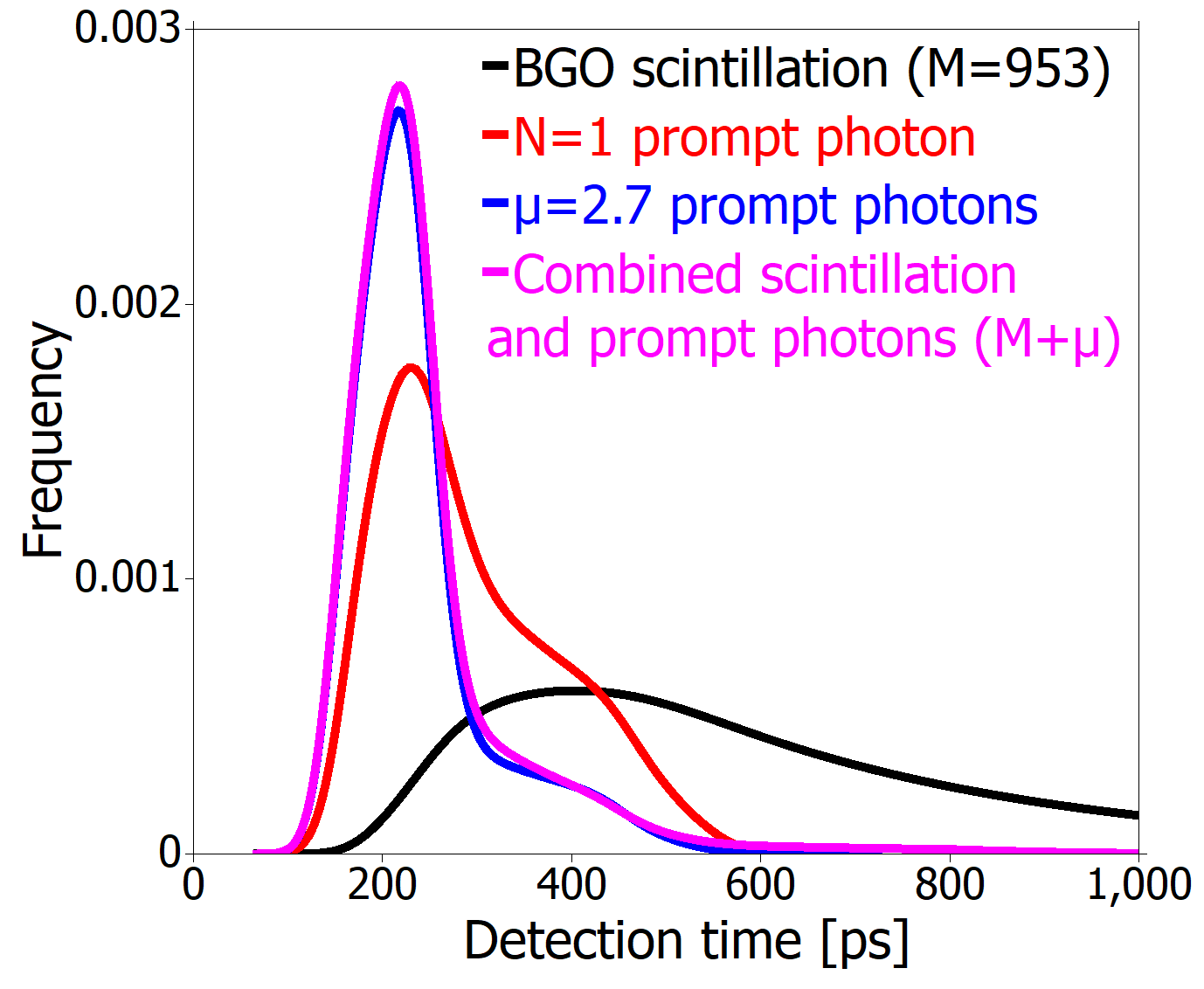}   
    \includegraphics[width=0.49\textwidth]{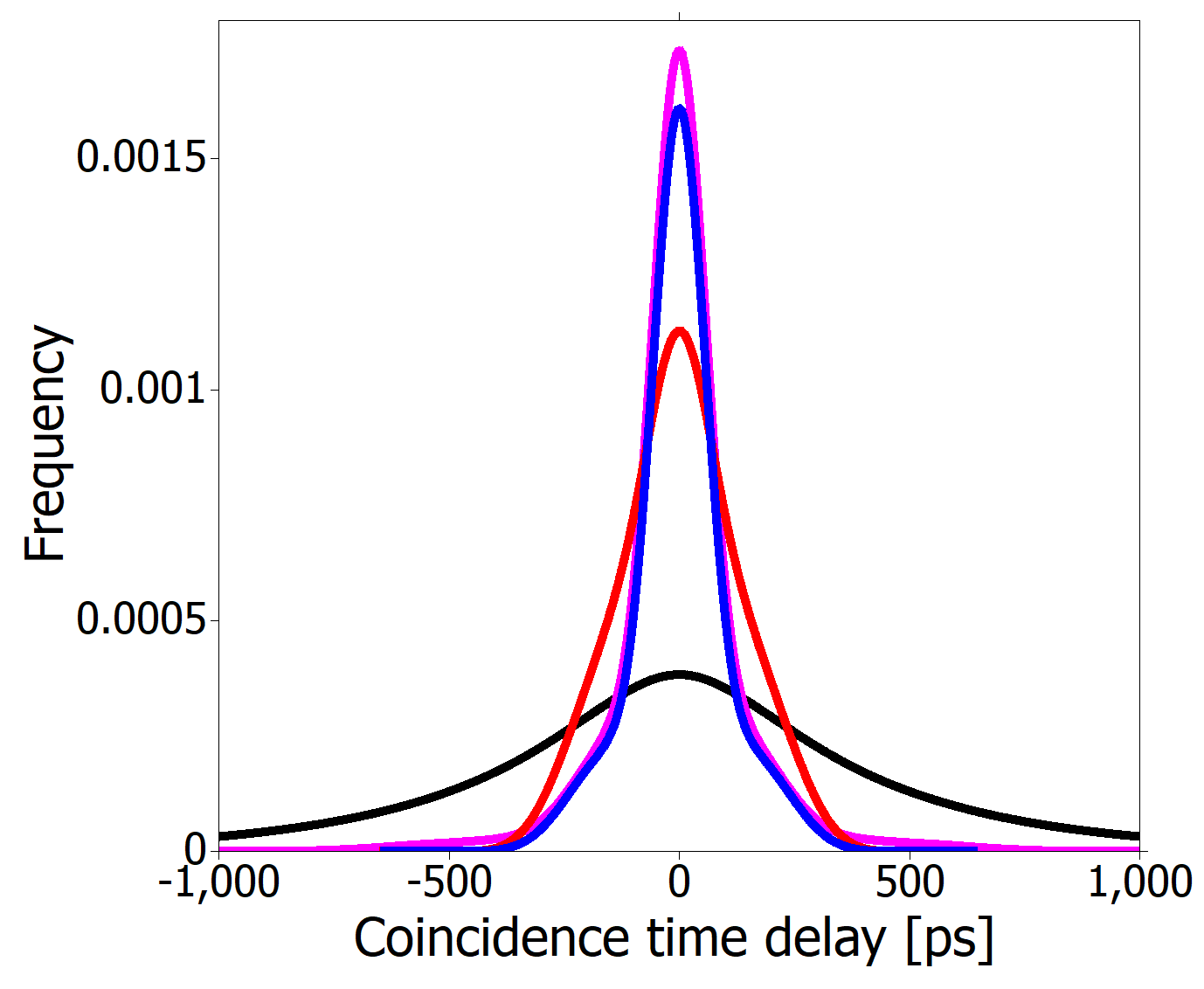} \\
\end{center}
\caption{Top: Detection time probability distribution of the first detected scintillation photon for different number of  M detected photons in linear scale for the first nanosecond (left) and logarithmic scale for the first 10~ns (right).
Bottom left: detection time probability distribution of the first photon considering only BGO scintillation (black), only one detected prompt photon (red), an average of $\mu=2.7$ detected prompt photons with underlying Poisson fluctuations (blue), and combined BGO scintillation with Cherenkov photons (magenta).
Bottom right: Coincidence time delay probability distribution considering the same cases.}
\label{fig:fist-photon}
\end{figure}

The scintillation production time profile is modeled as a sum of bi-exponential functions defined in equation~\ref{eq:scint}.
\begin{equation}
\label{eq:scint}
    f_{\text{Scint}}(t) = \sum_{i} \frac{e^{\left(-t/\tau_{d,i} \right)} - e^{\left(-t/\tau_{r} \right)}}{\tau_{d,i} - \tau_r} \cdot R_i
\end{equation}
The distribution is truncated after 5-50~ns (depending on the material) and the truncated values (calculated at least up to 3 times the slowest decay time) added to the last valid bin.
Since late photons barely contribute to the timing resolution, this truncation speeds up the calculation time by orders of magnitude without impacting the results if the truncation is carefully adjusted based on the decay time and number of detected photons. 
The distribution is normalized and convolved with the light propagation and the photodetector response to calculate the detection time probability of a single (random) scintillation photon.
We consider a total of M scintillation photons detected and assume that each of them is independent but follows the same probability density function $PDF_{scint}$.
Based on the cumulative distribution function $F(t)$ outlined in equation~\ref{eq:CDF} ~\cite{david1981order,Seifert_2012_PMB}, we can then calculate the detection time distribution of the first photon $PDF_{scint,first,M}$.
%
%Having in total M scintillation photons detected and assuming each of them are independent, but following the same probability density function $PDF_{scint}$, we can calculate the detection time distribution of the first photon $PDF_{scint,first,M}$ based on the cumulative distribution function $F(t)$ outlined in equation~\ref{eq:CDF} ~\cite{david1981order,Seifert_2012_PMB}.
%
\begin{equation}
\begin{split}
& F(t) = \int_{0}^t PDF_{scint}(\tau) ~d\tau \\
& F_{first}(t) = 1-\left[1-F(t) \right]^M \\
& PDF_{scint,first,M}(t) = \frac{d}{dt} F_{first}(t) \\
& PDF_{scint,first,M}(t) = M \cdot \left[1-F(t) \right]^{M-1} \cdot PDF_{scint}(t)
\end{split}
\label{eq:CDF}
\end{equation}
For the more general case considering the m-th photon, the arrival time is calculated according to equation~\ref{eq:CDF-m}.
\begin{equation}
\begin{split}
PDF_{scint,m-th,M}(t) &= \frac{M!}{(m-1)!(M-m)!} \cdot \left[F(t) \right]^{m-1} \\ 
&\quad  \cdot \left[1-F(t) \right]^{M-m} \cdot PDF_{scint}(t)
\end{split}
\label{eq:CDF-m}
\end{equation}
The photon detection time distribution for different number of $M$ for BGO with a 20~mm thick crystal (after averaging over all DOIs, as discussed in section~\ref{sec:DOI}) is displayed in the top of figure~\ref{fig:fist-photon}.
\subsection{Prompt photons, fluctuations and merged probability}
The convolution of any function with a Dirac-delta function gives the original function.
Hence, the detection time of a prompt photon is solely the light transport convolved with the SPTR, as shown in figure~\ref{fig:SPTR}.
For a total number of N prompt photons detected, the detection time of the first photon $PDF_{Cher,first,N}$ can be calculated analog to the case of scintillation light outlined in equation~\ref{eq:CDF}.
We assume that the number of detected prompt photons follows a Poison-like distribution~\cite{Kratochwil_CPB-2025}, so the detection time of the first prompt photon with a mean number $\mu$ is calculated considering 1, 2, ... ($3\cdot\mu+8$) prompt photons.
The cutoff at $3\cdot\mu+8$ is arbitrary to ensure even small probabilities for any $\mu$ are included.
The detection time of the first prompt photon is thus the weighted average over all probabilities according to equation~\ref{eq:Cher}.
\begin{equation}
\begin{split}
& PDF_{Cher,first,\mu} = \sum_{N=1}^{3\cdot\mu + 8} P_N \cdot PDF_{Cher,first,N} \\
& P_N = \frac{\mu ^N \cdot \text{e}^{-\mu}}{N!}; ~~~~~~~\sum_{N=0}^{3\cdot\mu + 8} P_N = 1
\end{split}
    \label{eq:Cher}
\end{equation}
The probability for $P_{N=0}$ is manually added to the last bin in a similar fashion as done for the truncation of scintillation light.
This ensures the correct normalization of the resulting PDF.
The PDF for $N=1$ and $\mu=2.7$ are displayed on the bottom left of figure~\ref{fig:fist-photon}.
The PDF of the first detected optical photon is calculated according to equation~\ref{eq:PDF-first}, using the definition of the cumulative distribution from equation~\ref{eq:CDF}.
\begin{equation}
\begin{split}
PDF_{first,joint}(t) &= PDF_{Cher,first,\mu}(t) \cdot \left[1 - F_{scint}(t) \right]\\ &+ PDF_{scint,first,M}(t) \cdot \left[1-F_{Cher}(t) \right]
\end{split}
    \label{eq:PDF-first}
\end{equation}
As alternative to equation~\ref{eq:PDF-first}, we can construct the average photon distribution~\cite{Gundacker_2016_PMB} by weighting the respective detection yield according to equation~\ref{eq:average-PDF} and examine the photon detection time probability of the first photon, considering in total (M+$\mu$) detected photons using equation~\ref{eq:CDF}.
\begin{equation}
    PDF_{avg}(t) = \frac{M \cdot PDF_{scint}(t)}{M+\mu} + \frac{\mu\cdot PDF_{Cher}(t)}{M+\mu} 
    \label{eq:average-PDF}
\end{equation}
Although this average PDF does not explicitly consider Cherenkov fluctuations, for the purpose of looking at the detection time of individual photons, fluctuations identical to the Poisson distribution with mean number $\mu$ are intrinsically present.
This originates from the Poisson limit theorem, and identical results are achieved in this manuscript for both approaches considering the first detected photon.
%
%Although this does not explicitly consider Cherenkov fluctuations, intrinsically they are present and identical results are achieved for the first detected photon. 
%
This is consistent to the findings reported in~\cite{Loignon-Houle_2023_TRPMS}.
\subsection{Fisher information and CRLB}
The Fisher information~\cite{Fisher1922-FISOTM} and the Cramér-Rao Lower Bound~\cite{cramer1946mathematical,rao1945information,rao1948large,Nielsen_2013-arxiv} are fundamental concepts in statistical estimation theory that characterize the information content and the achievable precision of an unbiased estimator, based on the underlying probability distribution of the observed data.
In the context of this manuscript, the detection time (or the time of the gamma-ray at DOI = 0) is the random variable, and the probability distribution describes to the detection time distribution.
Essentially, the CRLB states that the variance of any unbiased timing estimator $\hat{\theta}$ (e.g., first photon, second photon, average of multiple photons, ...) cannot be smaller than the variance predicted by the CRLB.
For a probability density function $f(t;\theta)$, the Fisher information $I(\theta)$- i.e, the information that the distribution carries about the unknown parameter $\theta$ (e.g., the true arrival time), is defined according to equation~\ref{eq:fisher}.
The CRLB for $i$ detected photons, converted to standard deviation times 3.33 to be consistent with the FWHM for a detector in coincidence, is the inverse square root of the product between the number of photons and the Fisher information.
\begin{equation}
\begin{split}
& I(\theta) = \int \left(\frac{\partial f(t;\theta)} {\partial \theta} \right)^2 \frac{1}{f(t;\theta)} ~dt \\
& \text{Var}(\hat{\theta}) \ge \frac{1}{i \cdot I(\theta)} \\
& CTR^{\text{CRLB}}_{\text{FWHM}} = 3.33 \cdot \sqrt{\frac{1}{i \cdot I(\theta)}}
    \label{eq:fisher}
\end{split}
\end{equation}
In the context of the CRLB, using the PDF of an average photon from equation~\ref{eq:average-PDF} gives overoptimistic~\cite{Loignon-Houle_2023_TRPMS} values.
Instead, fluctuations need to be added by manually constructing multiple PDFs with probabilities $P_N$, depending on the number of detected prompt photons according to equation~\ref{eq:CRLB-fluct}
\begin{equation}
    \begin{split}
& PDF_{fluct}(t) = \sum_N PDF_{fluct,N}(t) = \\ &\sum_{N=0}^{3\cdot\mu+8} P_N \cdot \left(\frac{M\cdot PDF_{scint}(t)}{M+N}  \right. + \left. \frac{N\cdot PDF_{Cher}(t)}{M+N}  \right)
        \label{eq:CRLB-fluct}
    \end{split}
\end{equation}
The Fisher information is then calculated for each $N$ individually according to~\ref{eq:Fisher-split}.
In contrast to~\cite{Loignon-Houle_2023_TRPMS}, we always treat the PDFs of scintillation and Cherenkov photons together, as separating them does not fully account for the combined statistical variance. %can lead to overoptimistic results.
By analogy with the no-hair theorem~\cite{Ruffini1971_PT}, photons have 'no hair'—they do not carry explicit information about their production mechanism and should therefore be treated collectively.
The calculation of the CRLB in equation~\ref{eq:Fisher-split} also considers different total number of detected photons, although for the approximation of $(M+N) = M$ the simplification of $CRLB = 3.33 \cdot \sqrt{1/(M \cdot I)}$ can be done.
\begin{equation}
    \begin{split}
%& PDF_{fluct,N}(t) = \frac{M\cdot PDF_{scint}(t)}{M+N}   + \frac{N\cdot PDF_{Cher}(t)}{M+N}  \\
& I_N(t_0) = \int \left(\frac{d PDF_{fluct,N}(t)}{dt}^2  \right) \cdot \frac{1}{PDF_{fluct,N}(t)} ~dt \\
& I(t_0) = \sum_N P_N \cdot I_N \\
& CRLB = 3.33 \cdot \sqrt{\sum_N \frac{P_N}{ (M+N) \cdot I_N(t_0)}}
        \label{eq:Fisher-split}
    \end{split}
\end{equation}
The truncation for scintillation and Cherenkov light adds a second peak to the PDF at the very end with a high $d PDF/dt$ without physical reason.
To avoid this, the integration for $I_N$ was stopped after 90\% of the total timescale.

%Because of the truncation for scintillation and Cherenkov light, which adds a second peak to the PDF at the very end (with a unnaturally high $d PDF/dt$) without physical reason, the integration for $I_N$ was only done for 90\% of the total timescale.
%
Also, we avoid numeric instabilities by dividing over too small probability values by only integrating values with $PDF > 1e-12$.
Within the uncertainty, no differences are observed changing the the integration cutoff to 60\%, or decreasing the integration threshold to $1e-9$.

We note that the Fisher information is generally not additive for mixtures of different distributions.
In our case, however, fluctuations occur on an event-by-event basis, and event classification is feasible~\cite{Kratochwil_2020_PMB}.
%
%Nonetheless, since our approach adapts a statistical concept originating from a field quite distant from instrumentation and detector physics, we acknowledge that the underlying approximations may not be fully mathematically justified in this context.
%
%It therefore remains uncertain whether the reported values are slightly optimistic or whether the assumptions hold strictly, given that the events are treated as independent.
%
%
\subsection{Depth-of-interaction averaging}
\label{sec:DOI}
Throughout the explanations in the text beginning with section~\ref{sec:light-propagation} only a single DOI is considered.
All calculations are repeated for each DOI in steps of $dDOI = 0.5$~mm.

%So far, all the calculations are done for a single DOI and all the steps beginning with section~\ref{sec:light-propagation} are repeated for all DOIs in steps of $dDOI = 0.5$~mm.
%
The gamma-ray interaction probability is calculated based on the attenuation coefficient of the material and serves as relative weighting for equation~\ref{eq:PDF-first} to obtain the average detection probability of the first photon indicated in equation~\ref{eq:DOI-weighting}.
\begin{equation}
    \begin{split}
& P_{att}(DOI) = e^{-\frac{(DOI-dDOI/2)}{\tau_{att}}} - e^{-\frac{(DOI+dDOI/2)}{\tau_{att}}} \\
& \hat{P}_{att}(DOI) =  \frac{P_{att}(DOI)}{\sum_{DOI}P_{att}(DOI)} \\
& PDF_{DOI}(t) = \sum_{DOI} \hat{P}_{att}(DOI) \cdot PDF_{first}(t,DOI)
        \label{eq:DOI-weighting}
    \end{split}
\end{equation}
The comparison of DOI average photon detection time and individual DOIs is shown in figure~\ref{fig:SPTR}.
Since the PDF is shifted for each DOI, the DOI-induced time bias is intrinsically included.

The Fisher information, however, is invariant to any time translation and therefore the DOI-induced time bias need to be added manually~\cite{Toussaint_2019_PMB}.
The variance of a biased distribution (with a bias shifted from its mean $\mu$ to $\mu +D$) is calculated as the sum of the distribution variance plus the squared bias ($D^2$).
The DOI bias $D$ is the time difference between any arbitrary (but fixed) reference time and the detection of optical photons.
A simple reference system may be the time when the gamma-ray enters the crystal at $DOI=0$, in which case the offset is $\Delta t(DOI)$.
Any constant (like the term $L\cdot n /c$) can be removed, and by placing the reference system to the time the gamma-ray reaches (or would reach) the coordinate $x_0\approx L/2 - \mathcal{O}{(L^2/\tau_{att})}$ in the crystal, the overall DOI-bias is expected to be minimized.

The DOI-biased variance (corresponding to the Fisher information at a fixed DOI times the number of total photons $M$) is calculated by combining the Fisher information with the DOI bias and average the sum for all depths.
Combining the summation over all number of prompt photons $N$ from equation ~\ref{eq:Fisher-split} with all DOIs, the CRLB with DOI bias can be calculated as outlined in equation~\ref{eq:CRLB-DOI}.
The unbiased CRLB with DOI information available simplifies by setting $\Delta t_{min}(DOI) = 0$. 
\begin{strip}
\begin{equation}
    \begin{split}
& \Delta t (DOI) = \frac{DOI}{c} + \frac{n \cdot (L-DOI)}{c} \rightarrow \frac{DOI \cdot (1-n) \cancel{+ L \cdot n}}{c} \\
& x_0 = \tau_{att} - \frac{L \cdot e^{-L/\tau_{att}}} {1-e^{-L/\tau_{att}}}; ~~~~~~~~~~~~~~~ \Delta t_{min} (DOI) = \frac{(DOI-x_0)\cdot (1-n)}{c}\\
& \text{Var}_{DOI,~biased}(\hat{\theta}) \ge \sum_{DOI} \hat{P}_{att}(DOI) \cdot \left(\frac{1}{I(t_0,DOI) \cdot M} + \Delta t^2_{min}(DOI) \right)\\
%& \text{Var}_{DOI,~unbiased}(\hat{\theta}) \ge \sum_{DOI} \frac{\hat{P}_{att}(DOI)}{I(t_0,DOI) \cdot N} \\
& CRLB_{biased} = 3.33 \cdot \sqrt{\sum_{DOI}\hat{P}_{att}(DOI) \cdot \left( \Delta t^2_{min}(DOI) + \sum_N \frac{P_N}{(M+N) \cdot I_N(DOI)}\right)} \\
%& CRLB_{unbiased} = 3.33 \cdot \sqrt{\sum_{DOI}\sum_N   \frac{\hat{P}_{att}(DOI) \cdot P_N}{(M+N) \cdot I_N(DOI)}} \\
        \label{eq:CRLB-DOI}
    \end{split}
\end{equation}
\end{strip}
\subsection{Probability time delay distribution and timing metrics}
The detection time probability distribution from equation~\ref{eq:DOI-weighting} is used to calculate the probability distribution of the time delay between two identical detectors.
This is achieved by auto-convolution of $PDF_{DOI}(t)$ with its reflected version according to equation~\ref{eq:self-convolution}.
\begin{equation}
    PDF_{\Delta T}(t) = \int_{-\infty}^{\infty} PDF_{DOI}(\tau) \cdot PDF_{DOI}(\tau+t) ~d\tau
    \label{eq:self-convolution}
\end{equation}
This distribution corresponds to a typical coincidence time delay histogram which is normalized to the total number of events being $\int PDF_{\Delta T}(t)~dt =1$.
The CTR distribution and the parent photon detection time distribution are shown in the bottom of figure~\ref{fig:fist-photon}.
The FWHM is extracted from $PDF_{\Delta T}(t)$.
Furthermore, the SNR equivalent CTR~\cite{Nuyts-TMI-2023,Kratochwil_TRPMS_2025_dual} is calculated according to equation~\ref{eq:CTR-SNR}.

\begin{equation}
CTR(SNR) = \frac{2 \cdot \sqrt{2 \text{ln}(2)}}{2 \cdot \sqrt{\pi}} \cdot \frac{dt}{\int PDF^2_{\Delta T}(t) ~dt}
    \label{eq:CTR-SNR}
\end{equation}
Lastly, the standard deviation (times 2.355 to convert into FWHM) of the distribution is extracted to have a reasonable comparison with the CRLB.
The FWHM is the least sensitive to tails or outliers in the time delay distribution, the CTR (SNR) is somewhat sensitive, and the standard deviation and the CRLB are the most sensitive to tails.
\section{Results and discussion}
\subsection{Model validation}
\begin{figure*}[h!] 
\begin{center}     
    \includegraphics[width=0.48\textwidth]{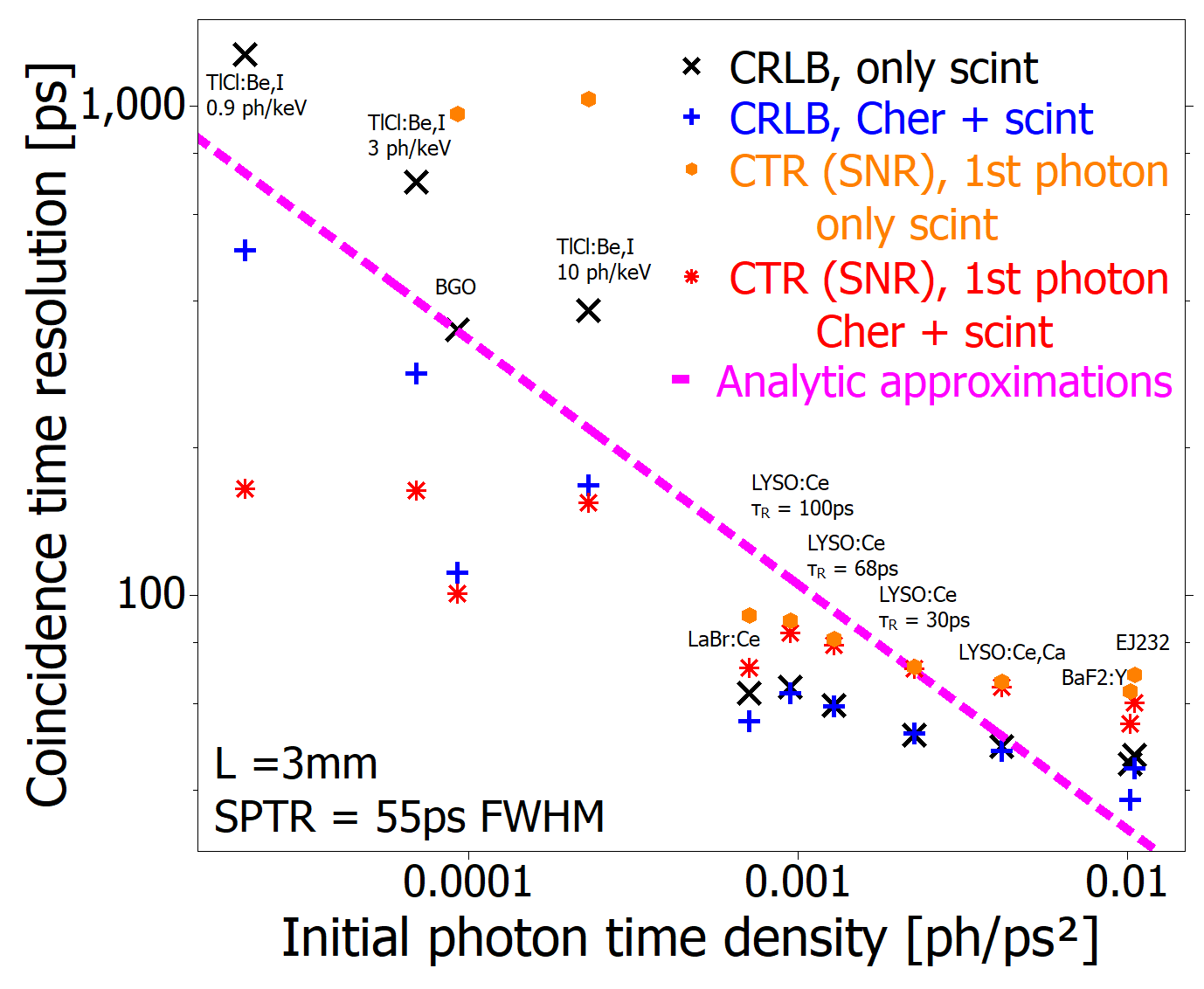} 
    \includegraphics[width=0.48\textwidth]{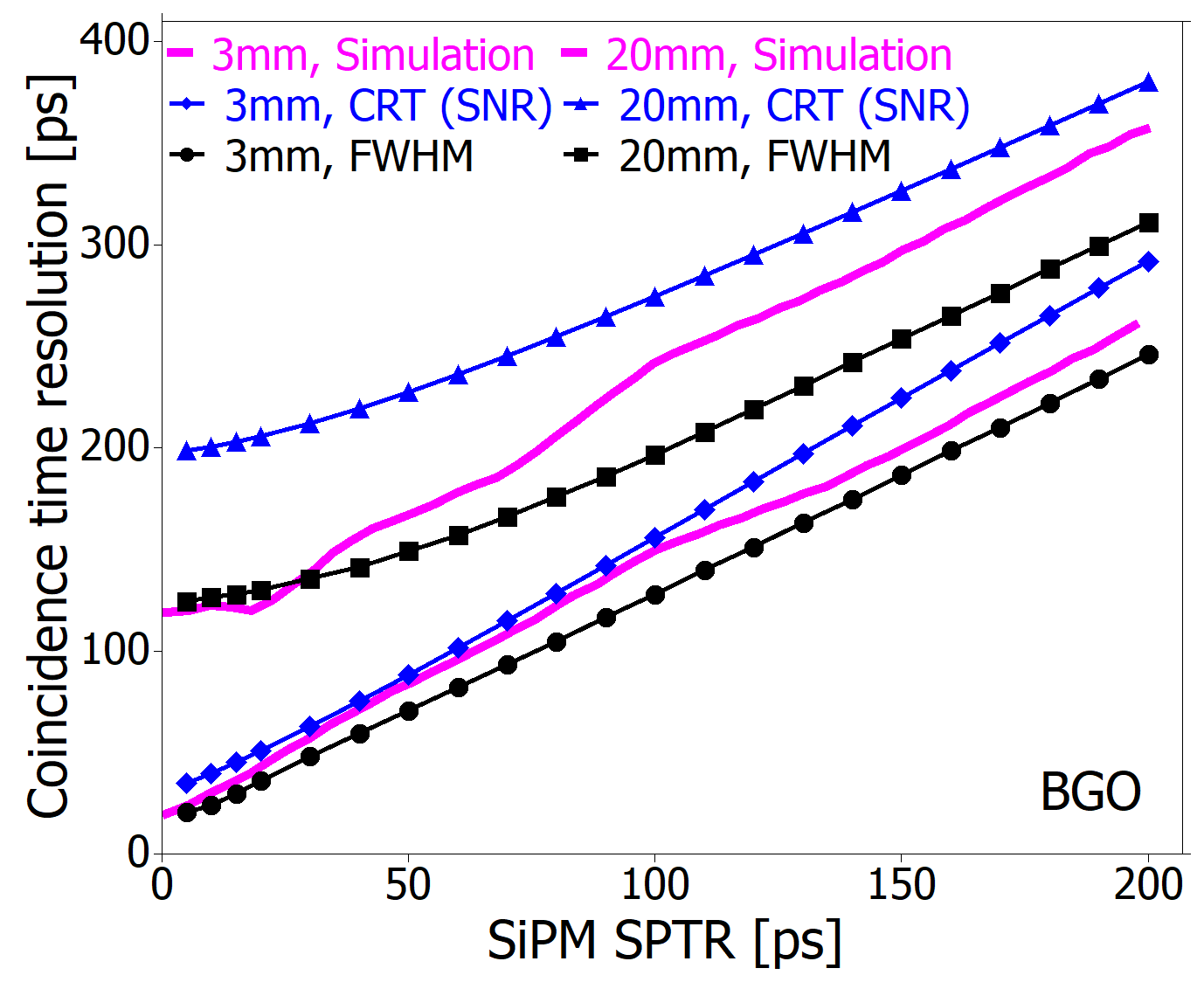} 
\end{center}
\caption{Left: Comparison of time resolution results of the analytic framework (this study) with approximations~\cite{VINOGRADOV_2018_NIMA} for different materials.
The input parameters of each material is specified in table~\ref{tab:crystals} and, if modified, indicated in the figure. 
The initial photon time density (IPTD) is calculated as outlined in~\cite{Gundacker_2020_PMB}.
The analytic approximation is 3.33 times the inverse square root of the IPTD.
The CRLB and CTR (SNR) are calculated for each material with and without the presence of prompt photons on top of scintillation.
Right: CTR comparison of Monte-Carlo simulations~\cite{Gundacker_2020_PMB} (FWHM) and our analytic framework for two different BGO geometries as function of the SiPM SPTR.}
\label{fig:validation}
\end{figure*}
We validate our model by comparing it with established analytic approximations from~\cite{VINOGRADOV_2018_NIMA,Gundacker_2020_PMB} for different crystal parameters.
The analytic approximation~\cite{VINOGRADOV_2018_NIMA} does not consider prompt emission or DOI-induced time bias / light transport, hence we only consider small (3~mm thick) crystals and show results with and without prompt photon emission. 
Additionally, we compare our calculations with Monte Carlo simulations~\cite{Gundacker_2020_PMB} of BGO for thin (3~mm) and thick (20~mm) crystals coupled to digital-like SiPMs with varying SPTR.
This dual approach supports the model's accuracy and broad applicability.

Figure~\ref{fig:validation}, left, shows the calculated CTR with and without prompt photons for various crystals and input parameters in terms of CRLB and CTR (SNR).
Our model matches very well for BGO (+3~ps, +0.8\%)  and LYSO:Ce,Ca (\text{-2~ps}, \text{-5\%}) considering the CRLB and only scintillation photons.
Our results are worse than the approximation for TlCl:Be,I (0.9~ph/keV: +542~ps, +75\%)  and better for LaBr:Ce (-62~ps, -49\%), and LYSO:Ce ($\tau_r = 68$~ps: -18~ps, -26\%), while again becoming worse for very fast scintillators like BaF2:Y (+12~ps, +37\%) and EJ232 (+15~ps, + 45\%).
We attribute this difference to a breakdown of assumptions used in the analytical approximations and described below~\cite{VINOGRADOV_2018_NIMA}:
TlCl:Be,I has a relatively low scintillation light yield and multiple decay times ranging from 3~ns to 300~ns, hence the resulting shape is not well described by a simple exponential decay component.
Further, the approximation~\cite{VINOGRADOV_2018_NIMA} performs a Taylor expansion which requires that $\sigma_{PTS}, SPTR << \tau_d$, which is no longer valid for BaF2:Y and EJ232.
The overall trend of better performance for common scintillators may be a combination of the aforementioned effects together with a higher information content provided by our assumed digital photodetectors instead of analog devices.

The addition of Cherenkov photons significantly improves the CTR for TlCl and BGO, while it has little effect for LYSO.
In ultra-fast materials, the relatively high Cherenkov photon yield again enhances the CTR.
The CTR (SNR) of the first photon is generally worse than the CRLB with the exception of TlCl:Be,I and BGO.
At first glance it is highly counterintuitive for CTR values to surpass the CRLB, but this can be explained by different sensitivities of the CTR metrics to the tails in the TOF distribution which have a strong impact for materials with low light yield and slow scintillation decay times.
We observe no degradation in timing performance due to Cherenkov light; in most cases its addition leads to clear improvement over using scintillation light alone.

Overall, our model successfully reproduces the trends in timing performance as a function of scintillation parameters~\cite{VINOGRADOV_2018_NIMA}.
Differences may be due to the limited prediction power due to the approximations (eg. Taylor expansion) mentioned before.
%Deviations from our analytic predictions are more plausibly due to oversimplifications and the limited predictive power of those approximations~\cite{VINOGRADOV_2018_NIMA}, rather than deficiencies in our model.

%
Figure~\ref{fig:validation}, right compares the calculated CTR of BGO for 3~mm and 20~mm thick crystals with Monte Carlo simulation results extracted from~\cite{Gundacker_2020_PMB}.
Noteworthy difference is the wrapping of the crystal with Teflon in the simulation, while our model considers a bare crystal with ESR opposite to the photodetector readout.
Despite this difference, our model accurately reproduces simulated data, especially for small SPTR values.
In contrary to the simulation results, our calculations appear smoother with increased SPTR.
The difference between FWHM and CTR (SNR) is less pronounced for the thin BGO crystal, indicating that the CTR distribution is mostly described by a single Gaussian distribution.
For perfect (=0~ps) SPTR, the calculated CTR is about 20~ps FWHM and 125~ps FWHM for the thin and thick BGO crystals, respectively.
This large difference is not only due to the DOI-induced time bias, but also an increased photon time spread due to longer travel time of optical photons before detection.
\subsection{Impact of crystal thickness on the CTR}
\begin{figure}[h!] 
\begin{center} 
    \includegraphics[width=0.95\textwidth]{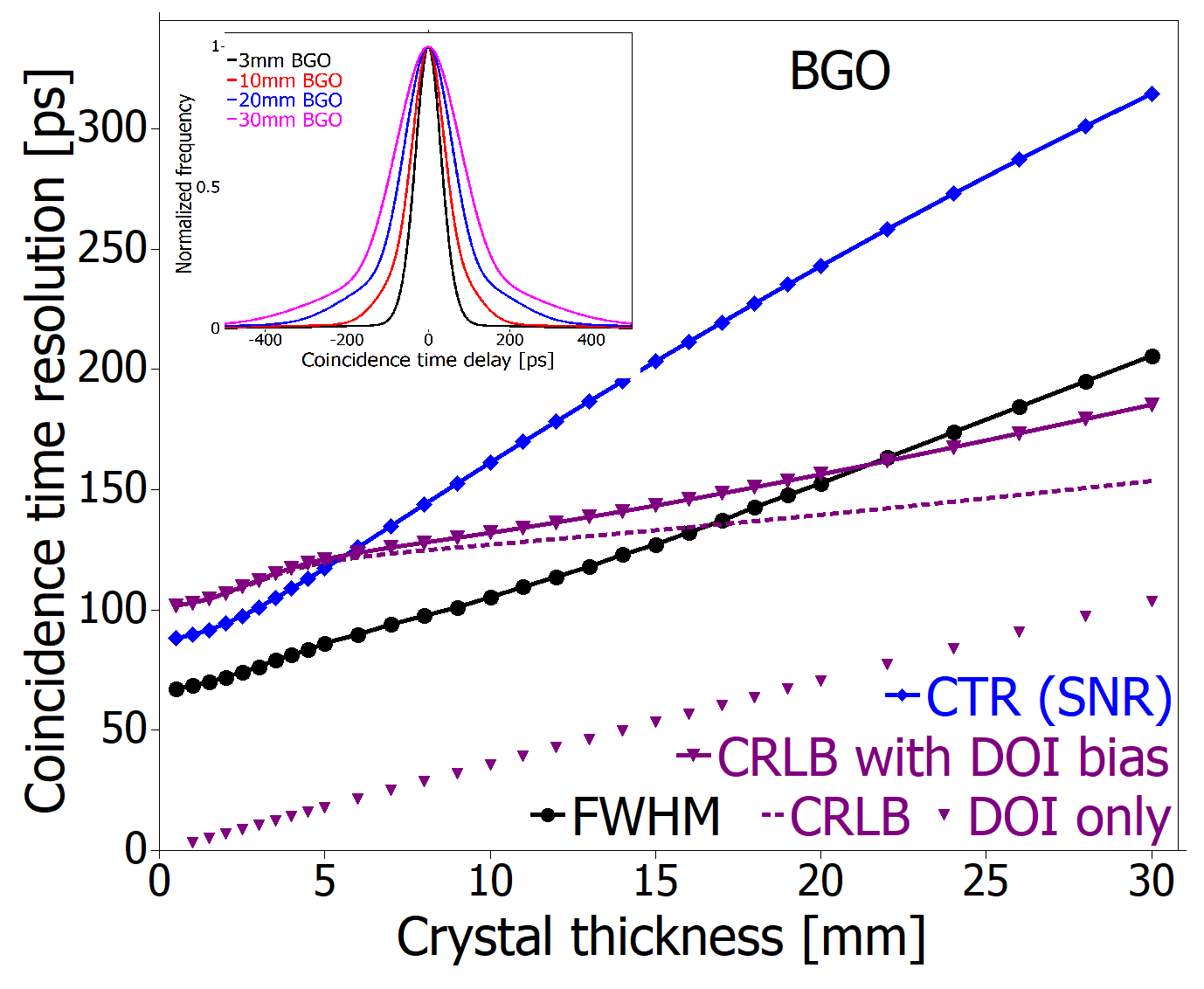}
\end{center}
\caption{Calculated time resolution of BGO for different crystal thickness.}
\label{fig:BGO-thickness}
\end{figure}
Figure~\ref{fig:BGO-thickness} shows the calculated time resolution of BGO (SPTR = 55~sp) with varying crystal thickness, ranging from 0.5~mm up to 30~mm.
The coincidence time delay histogram is shown as inset for different thicknesses.
For small thicknesses, the distribution is mostly Gaussian with small tails, which results in similar CTR (SNR) and FWHM. 
For increasing thickness, especially the CTR (SNR) becomes worse, while the FWHM is not that significantly impacted.
For small thickness (L $<$ 4~mm) the SiPM SPTR is the main contribution, while for larger thickness the increased photon time spread becomes the dominant contribution to the CTR (SNR). 
In the coincidence time delay distribution at least two regions are visible, with the core being Gaussian-shaped and the tails having an exponential behavior.
This is attributed to the interplay between the detection of Cherenkov and scintillation light, together with the increased light transport contribution.
The first photon may be produced at deep DOI and may thus travel almost twice the crystal thickness before being detected, which gives very pronounced tails. 

Initially, the CRLB is larger than the calculated FWHM or CTR (SNR).
This is due to different weighting of the tail. 
For reference, the standard deviation (not included in the figure) for 3~mm thick BGO using the first detected photon is about 450~ps and multifold larger than the FWHM or CTR (SNR).
Once the DOI bias on the CRLB becomes visible ($L~\approx~5$~mm), the slope of the CRLB with increasing thickness becomes less steep.
This may be as well explained by the interplay between SPTR and light transport contribution, since the latter becomes the dominant contribution on the CRLB.

The FWHM values are much better than experimental results with analog SiPMs~\cite{Kratochwil_IEEE_2021,Lee_2024_TRPMS}, since the SiPM signal and readout electronics contribution are not yet included in our calculations (see Fig. 12 in~\cite{Gundacker_2023_PMB}).
However, published experimental results~\cite{Cates_2019_PMB,Kratochwil_IEEE_2021} also report a broadening of the time delay histogram for thick crystals with more pronounced tails.

The fact that for thin crystals the CRLB ($\approx$ 110~ps) is much lower than the standard deviation ($\approx$~450~ps) when using only the timestamp of the first detected photon is encouraging.
It suggests that more advanced time estimation methods could further improve the achievable timing performance.
For example, incorporating the time difference between the first and second detected photon (analogous to the signal rise time in experiments with analog SiPMs~\cite{Kratochwil_2020_PMB}) may enhance the FWHM and CTR (SNR).
%
%
%The fact that the CRLB is with $\approx$ 110~ps (for thin crystals) much better than the respective standard deviation ($\approx~450~ps$) using only the timestamp of the first detected photon gives hope that a more advanced time estimation (eg. using the time difference between the first and second detected photon similar to the SiPM signal rise time in the experiment with analog SiPMs~\cite{Kratochwil_2020_PMB} as an additional information) further improve the achievable timing performance in terms of FWHM and CTR (SNR).
%
\begin{figure}[h!] 
\begin{center} 
    \includegraphics[width=0.95\textwidth]{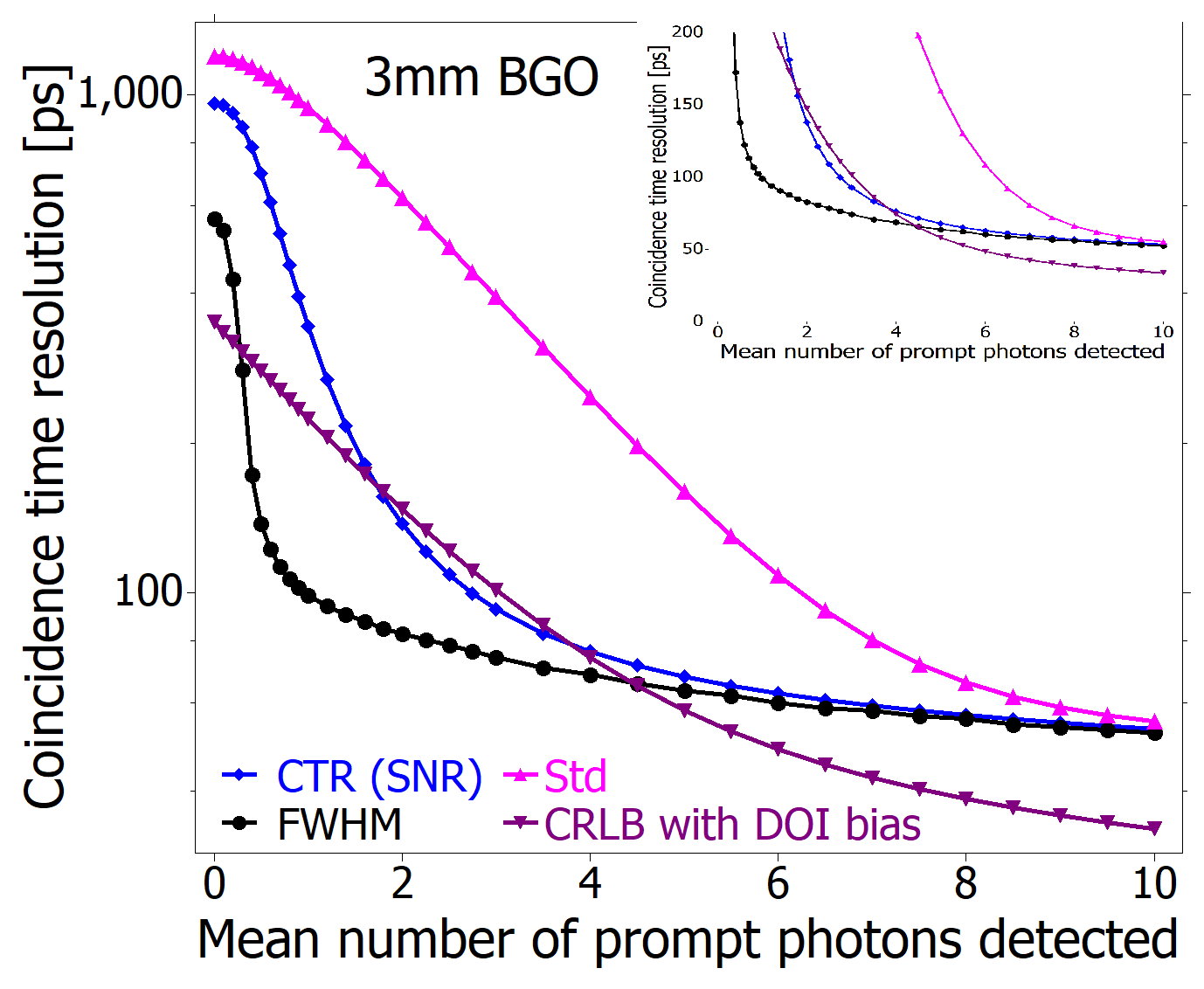} \\
        \includegraphics[width=0.95\textwidth]{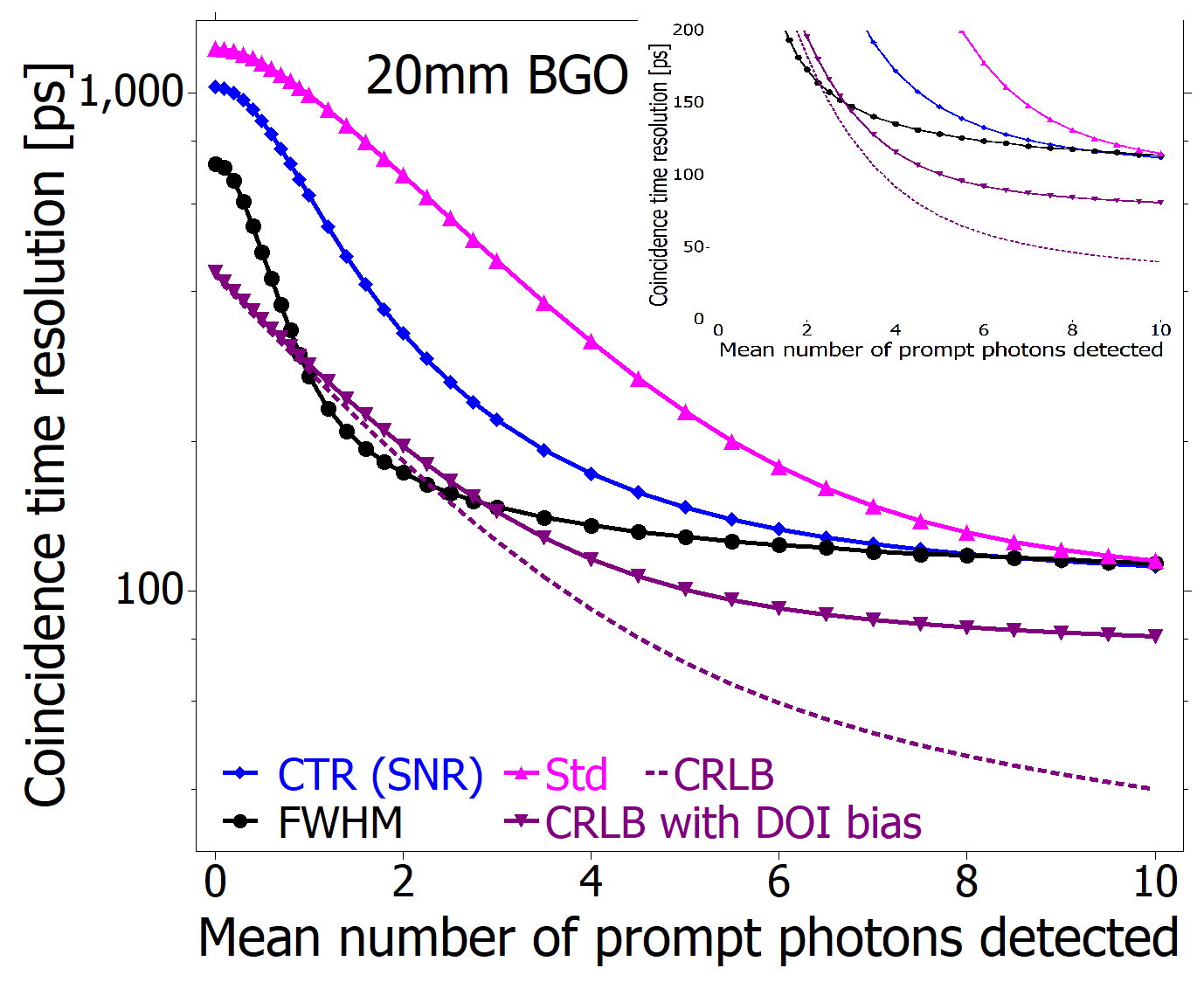} 
\end{center}
\caption{Calculated time resolution metrics as function of the mean number of detected prompt photons on top of the BGO scintillation emission for 3~mm (top) and 20~mm (bottom) thick BGO crystals. The inset shows the graphs in linear scale up to 200~ps.}
\label{fig:CRLB-mu}
\end{figure}
\begin{table*}[h!]
\centering
\caption{Input parameters and timing performance of tested scintillation and Cherenkov emitters}
\scalebox{0.75}{
\begin{threeparttable}
\begin{tabular}{l|c|c|c|c|c|c|c|c}
Quantity & Unit/Symbol & TlCl:Be,I & BGO & LaBr:Ce & LYSO:Ce & LYSO:Ce,Ca &  BaF$_2$:Y & EJ232 \\
\hline
Ref. index & n & 2.3 & 2.1 & 2.1 & 1.8 & 1.8 & 1.6 &  1.6 \\
Light transfer efficiency & LTE [\%] & 26.7 & 32.9 & 32.9 & 51.2 & 51.2 & 85.1 & 85.1\\
Scintillation yield & LY [ph/keV] & 0.9 (3 / 10 / 40)\tnote{c}  & 10.7 & 63 & 41.1 & 40 & 1.4 & 6\tnote{a}\\
Scintillation decay times & $\tau_{d,i}$ [ns] & (3, 50, 300) & (46, 365) & 25 & (21.5, 43.8) & (21, 46) & (0.1, 0.8) & (1.3, 7) \\
Abundance decay time & $R_i$ [\%] & (3, 33, 64) & (8, 92) &  100 & (13, 87) & (30, 70) & (22, 78) & (80, 20)\\
Scintillation rise time & $\tau_r$ [ps] & 20 & 8 & 200 & 68 & 10 & 4 & 30 \\
Cherenkov photons produced & N$_{Cher}$ & 14.5 & 18.3 & 25.6 & 10.6 & 10.6 & 25.1 & 15.4\tnote{b} \\
Weighted detection efficiency scintillation & PDE$_{scint}$ [\%] & 57 & 53 & 58 & 64 & 64 & 25 & 55\\
Weighted detection efficiency Cherenkov & PDE$_{Cher}$ [\%] & 48 & 45 & 46 & 41 & 41 & 33 & 44 \\
Photoelectric attenuation length & $\tau_{att}$ [mm] & 21.1 & 24.1 & $>$ 100 & 35.8 & 35.8 & $>$ 100 & $>$ 100\\ \hline
Effective scintillation decay time & $\tau_{d,eff}$ [ns]& 53 & 235 & 25 & 39 & 34 & 0.3 & 1.5\\
Number detected scintillation photons & M & 70 & 950 & 6140 & 6880 & 6700 & 150 & 1440\\
Number detected Cherenkov photons & $\mu$ & 1.9 & 2.7 & 3.9 & 2.2 & 2.2 &  7.0 & 5.8 \\
\hline \hline
CTR (SNR)\tnote{d}~~(20~mm crystal)& [ps] & 494 (450 / 375 / 260)\tnote{c}   & 244 & 134 & 131 & 113 & 99 & 108\\
FWHM\tnote{d}~~(20~mm crystal)& [ps] & 179 (183 / 189 / 195)\tnote{c}  & 153 & 137 & 134 & 115 & 101 & 110\\
CRLB with DOI bias\tnote{e}~~(20~mm crystal) & [ps] & 618 (352 / 218 / 155)\tnote{c}  & 156 & 96 & 90 & 81 & 68 & 72\\
%CTR (SNR)\tnote{d}~~(20~mm crystal)& ps & 493.5 (450.2 / 374.5 / 259.6)   & 243.8& 134.4 & 131.2 & 112.6 & 99.7 & 108.3\\
%FWHM\tnote{d}~~(20~mm crystal)& ps & 179.4 (182.5 / 188.6 / 194.7)  & 153.3 & 137.2 & 133.9 & 115.2 & 101.3 & 109.9\\
%3.33 x CRLB with DOI bias\tnote{e}~~(20~mm crystal) & ps & 617.9 (354.7 / 222.9 / 162.7)  & 162.7 & 110.4 & 96.7 & 88.3 & 75.5 & 79.8\\
\end{tabular}
\begin{tablenotes}
\item[a] Scaled down from 10~ph/keV to represent an energy deposition of 340~keV instead of 511~keV.
\item[b] For a hot-recoil electron of 340~keV.
\item[c] Tested for hypothetical improvements in light yield.
\item[d] Calculated for the first detected photon
\item[e] Considering all detected photons. The estimated DOI bias ranges between 38 ps (n=1.6) and 81~ps (n=2.3) in quadrature and coincidence.
\end{tablenotes}
\end{threeparttable}
}
\label{tab:crystals}
\end{table*}
\begin{figure*}[h!] 
\begin{center} 
    \includegraphics[width=0.48\textwidth]{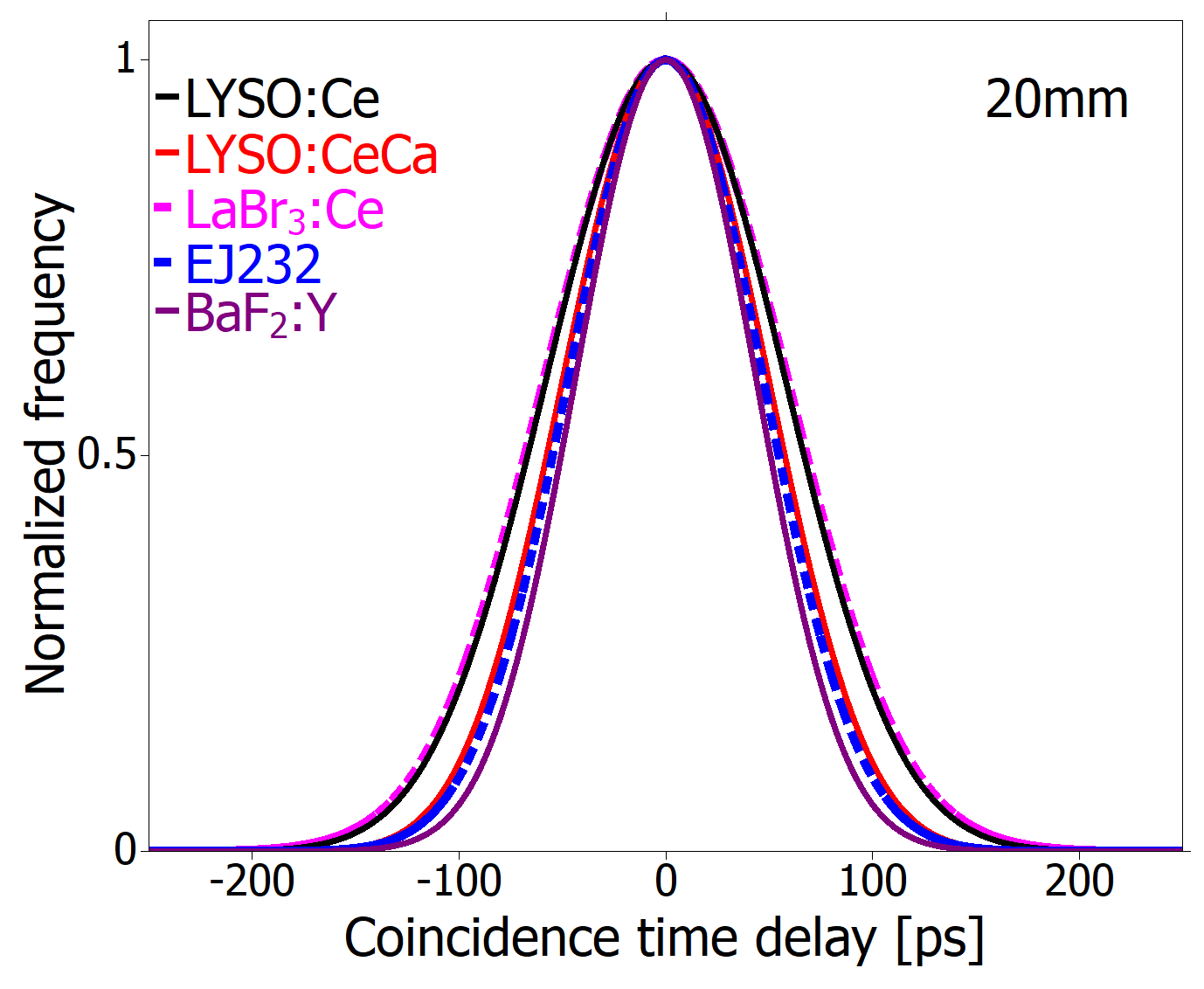}
        \includegraphics[width=0.48\textwidth]{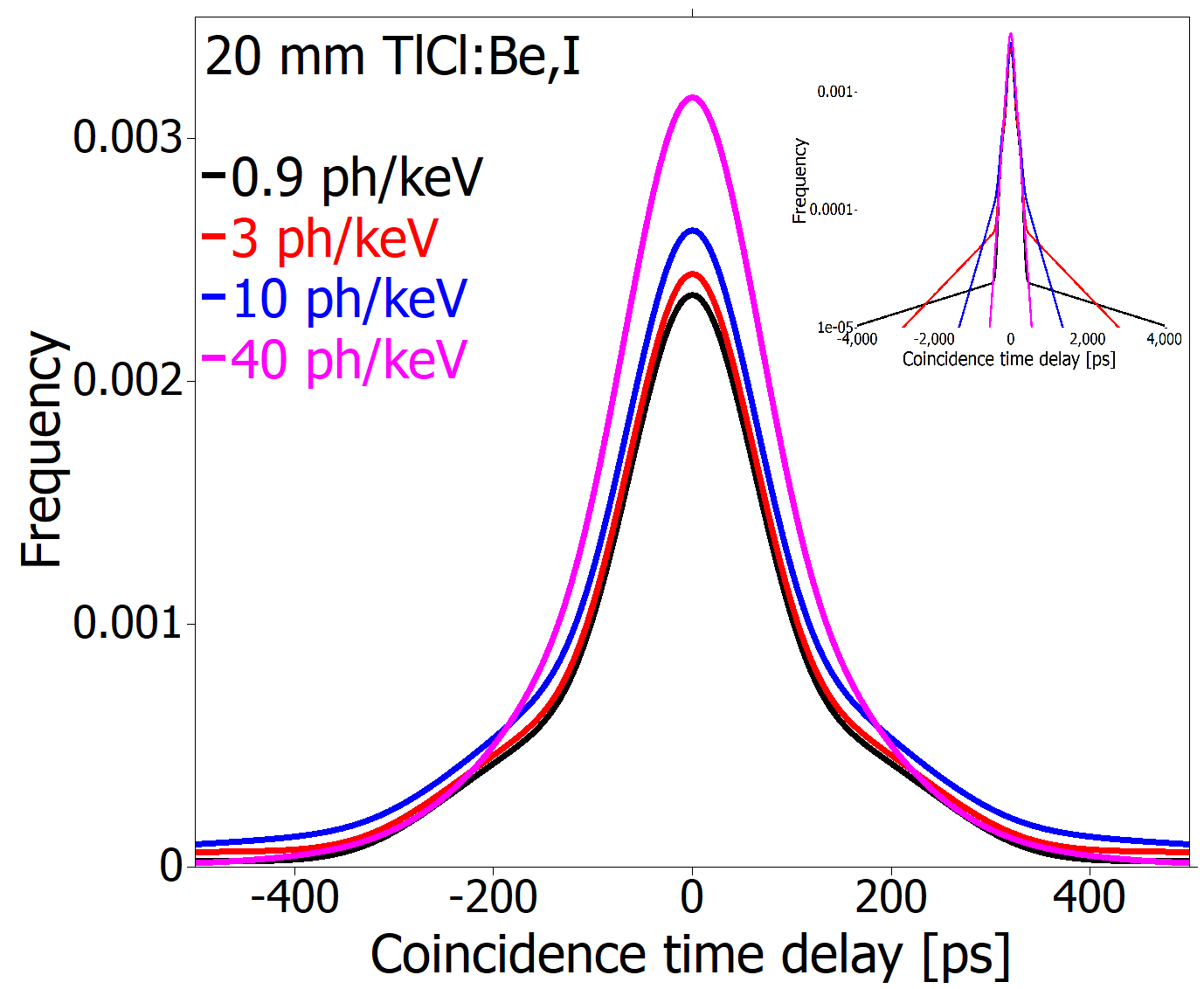} 
\end{center}
\caption{Left: Coincidence time delay distribution for selected scintillating crystals commonly used for fast-timing applications.
Right: Time delay distribution of TlCl:Be,I with reported scintillation yield of 0.9~ph/keV and hypothetical improvement up to LYSO yield of 40~ph/keV.
The inset is in logarithmic scale and up to $\pm$4~ns to better emphasize the tails in the distributions.}
\label{fig:CTR-materials}
\end{figure*}
\subsection{Timing performance metrics with prompt photons}
To further investigate the influence of prompt photons on the timing performance and in particular the shape of the TOF-distribution, we vary the mean number of detected prompt photons in BGO between 0 (only scintillation) up to 10.
This already accounts for the weighted detection efficiency and light transfer efficiency, hence 10 detected prompt photons correspond to 68~produced prompt photons.
Different timing metrics respond differently to the increase of detected prompt photons shown in figure~\ref{fig:CRLB-mu}.
The FWHM improves rapid for small Cherenkov photon numbers and already with an average of one detected prompt photons, CTR values below 100~ps FWHM (3~mm BGO) are obtained.
The SNR-equivalent CTR improves less significant, though once the probability to have no Cherenkov photons detected is sufficiently low ($\mu \approx 4$), it is similar to the FWHM of the TOF distribution.
The standard deviation (times 2.355) is very sensitive to tails and fluctuations on the number of detected prompt photons, thus only at high number of detected prompt photons ($\mu > 8$) it is compatible to FWHM and CTR (SNR) values.
Trends are similar for both crystal thicknesses, though the timing improvement (CTR (SNR), FWHM) requires more detected prompt photons.
The CRLB is worse than the FWHM for a wide range of mean number of detected prompt photons due to the complex TOF shapes.
Ultimately, once a high number of prompt photons are detected, the CRLB surpasses all other timing metrics confirming the central limit theorem of complex distributions to converge to a normal distribution.
Since the CRLB considers all detected photons, it is expected to eventually outperform other timing metrics.
The DOI-bias on the calculated CRLB is about 70~ps FWHM for the 20~mm thick BGO, hence only with DOI information this barrier can be surpassed.
For the extreme case of $\mu=10$, the CRLB doubles from 40~ps to 80~ps with DOI bias.
\subsection{Timing performance for different materials}
Calculations were repeated for selected scintillators and Cherenkov emitters commonly used for fast timing applications.
Table~\ref{tab:crystals} provides a detailed overview of the input parameters.
An attempt was made to keep literature values~\cite{Glodo-2005-TNS,Gundacker_2020_PMB, Gundacker_2021_PMB, Kratochwil_2023_PhD, Nadig_2023_PMB,Herweg_2023_TRPMS, Kratochwil_TRPMS_2024_TlCl,Arino-Estrada_2025_TRPMS} wherever possible, though assumptions were used when no information was available, or for conflicting literature results.
We assume photodetector properties similar to the NUV-MT technology~\cite{MT-Broadcom,Merzi_2023_JINST,Lee_2024_TRPMS} with SPTR values of 55~ps FWHM, but extended sensitivity down to the VUV (with about 25\% detection efficiency, inspired by~\cite{Datasheet-VUV-HPK}.
Cargille Meltmount (n$_M$ = 1.582) was modeled between the crystal and photodetector with a  transparency extended down to 200~nm.
Cherenkov photons are generated following the method outlined in~\cite{Kratochwil_TRPMS_2024_TlCl}.

Figure~\ref{fig:CTR-materials}, left shows the calculated coincidence time delay distribution for different 20~mm thick scintillators.
The timing performance (FWHM) ranges from 101~ps to 137~ps with similar SNR-equivalent CTR between 99~ps and 134~ps.
The best timing is typically achieved by triggering on later photons, which is well reflected in the calculated CRLB (with DOI bias) between 68~ps and 96~ps (see the last three lines in table~\ref{tab:crystals}).

Our framework allows to provide guidelines on crystal and doping schemes improvements such as TlCl:Be,I.
Different light yield values ranging from 0.9~ph/keV (current technology~\cite{Kratochwil_TRPMS_2024_TlCl}) all the way up to 10~ph/keV (similar to BGO) and 40~ph/keV (similar to LYSO) were modeled on the right of figure~\ref{fig:CTR-materials}.
The histogram is normalized so that the integral over the full time equals to 1.
With increased light yield, the probability to have events within a short timescale ($\pm$~200~ps) increases.
However, the FWHM slightly deteriorates from 179~ps to 195~ps, because events triggered by scintillation light are more likely to occur at small time delay differences.
Higher light yield leads to significant tail reduction, visible in the improvement in terms of CTR (SNR) and CRLB.
This means, R\&D efforts to increase the light yield of doped TlCl may lead to an counter-intuitive worsening of the timing performance when solely considering the FWHM.
Though, these efforts will pay off considering other metrics, since higher light yield (beside better energy resolution, etc) would lead to higher sensitivity and reduced tails.
\subsection{Photodetector improvements}
\begin{figure*}[h!] 
\begin{center} 
    \includegraphics[width=0.31\textwidth]{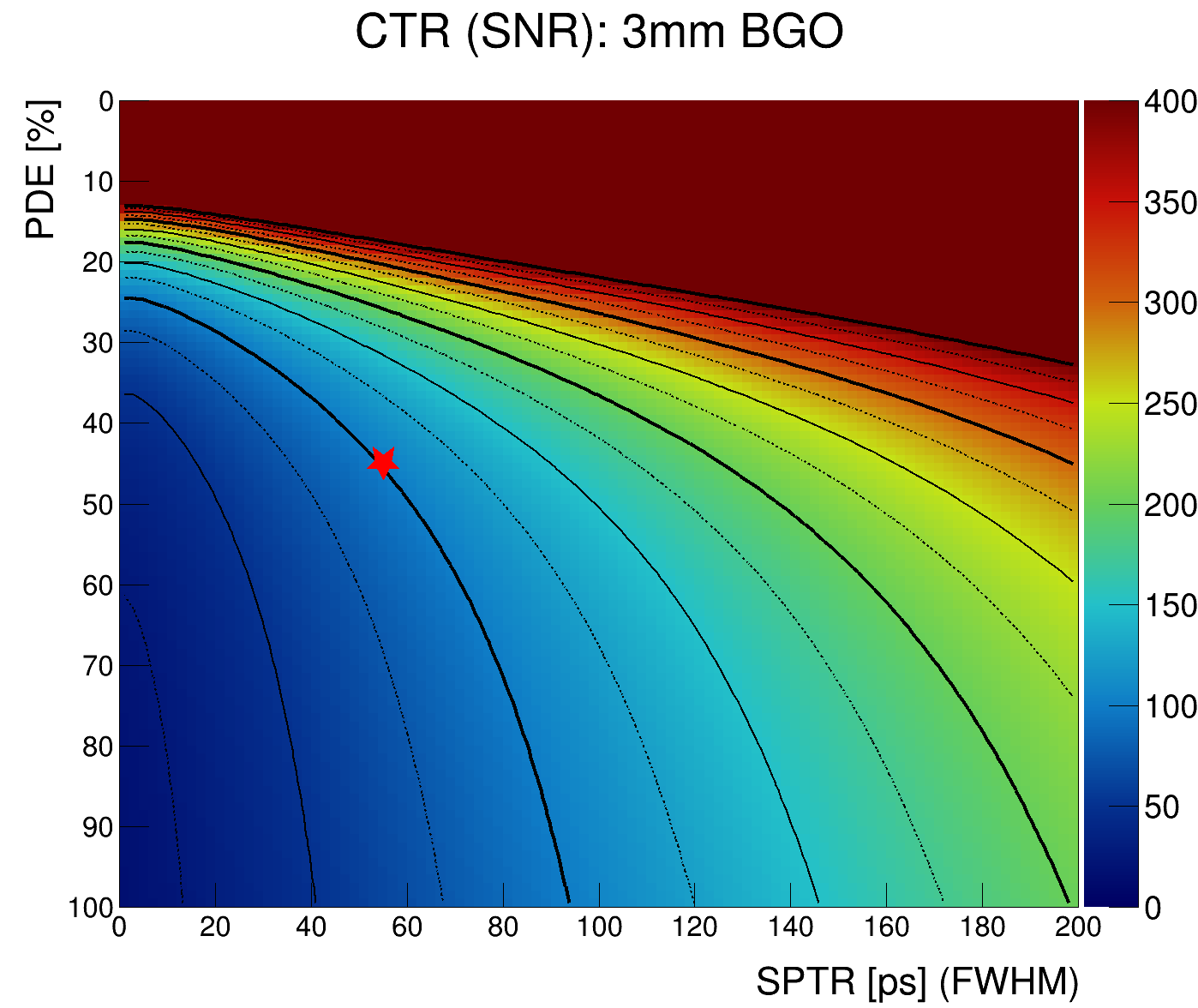}
    \includegraphics[width=0.31\textwidth]{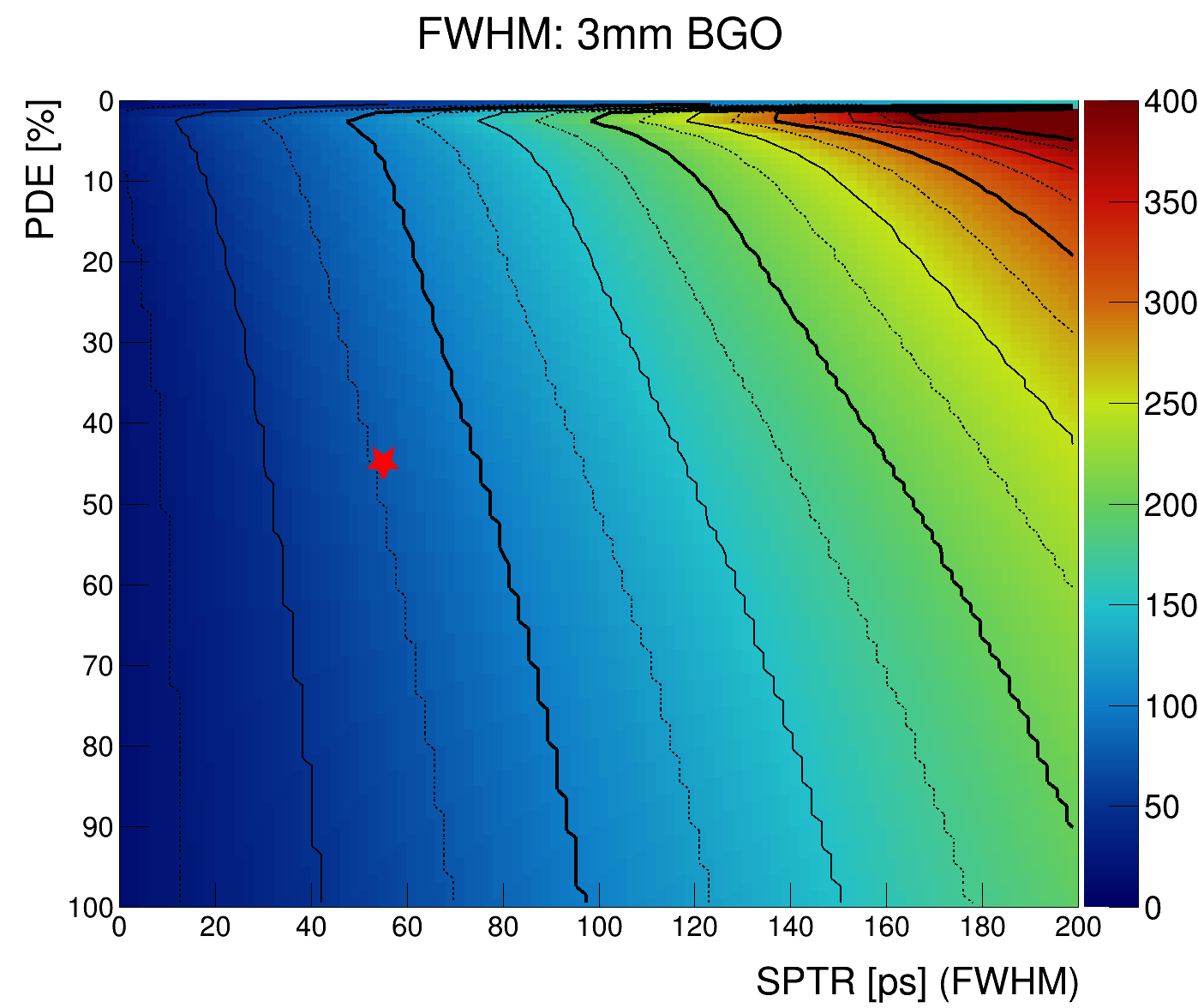} 
    \includegraphics[width=0.31\textwidth]{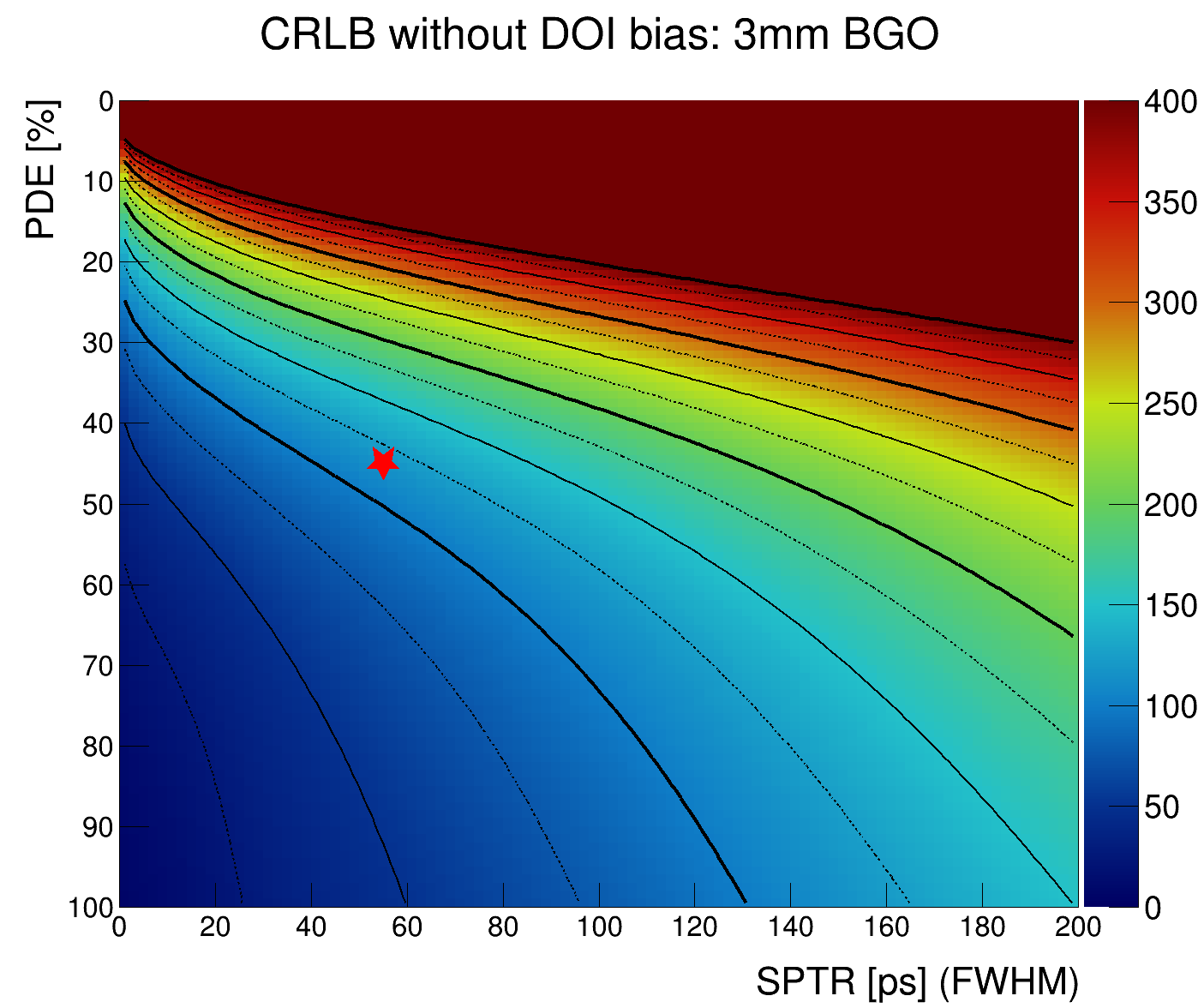} \\
                
        \includegraphics[width=0.31\textwidth]{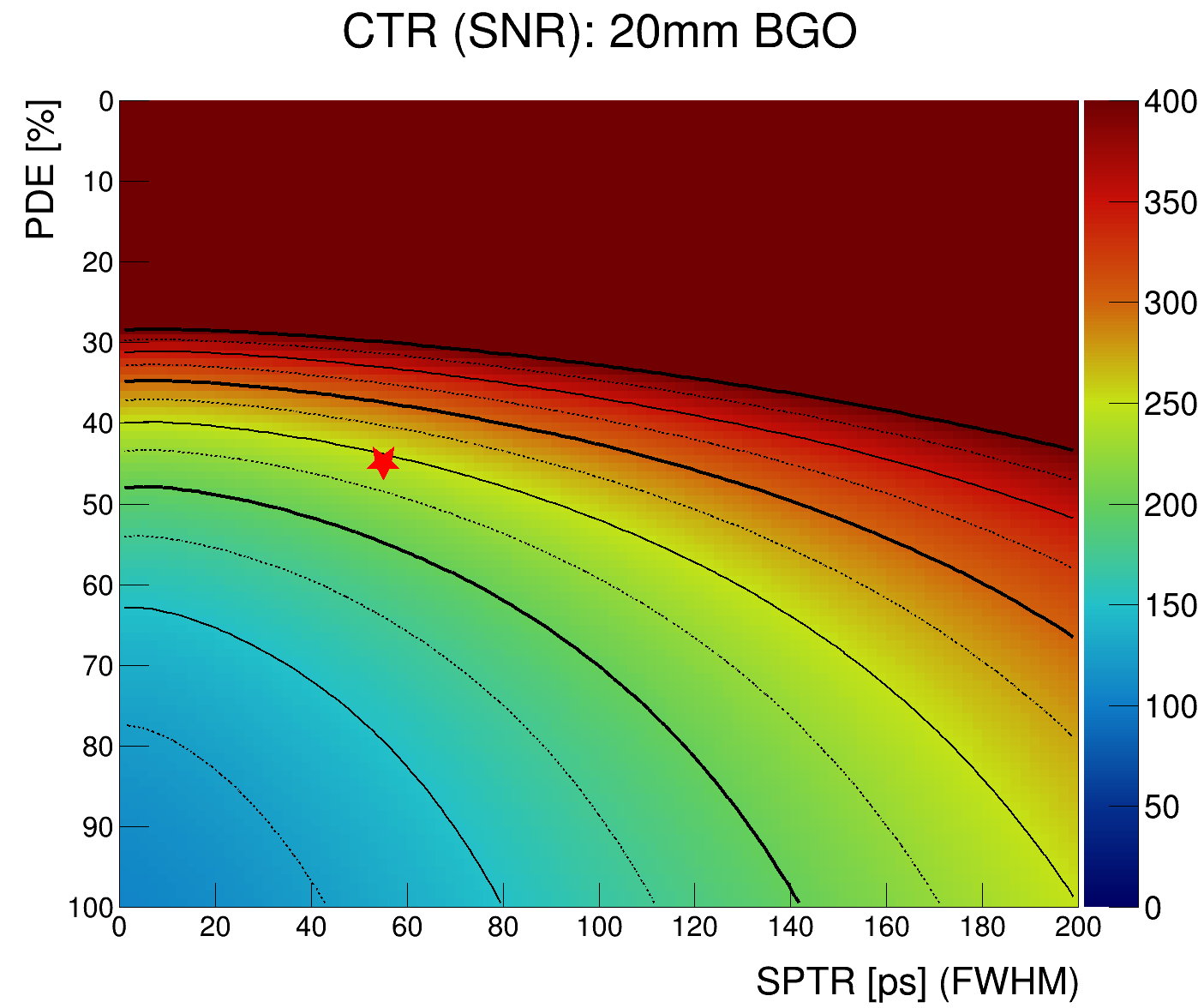}
        \includegraphics[width=0.31\textwidth]{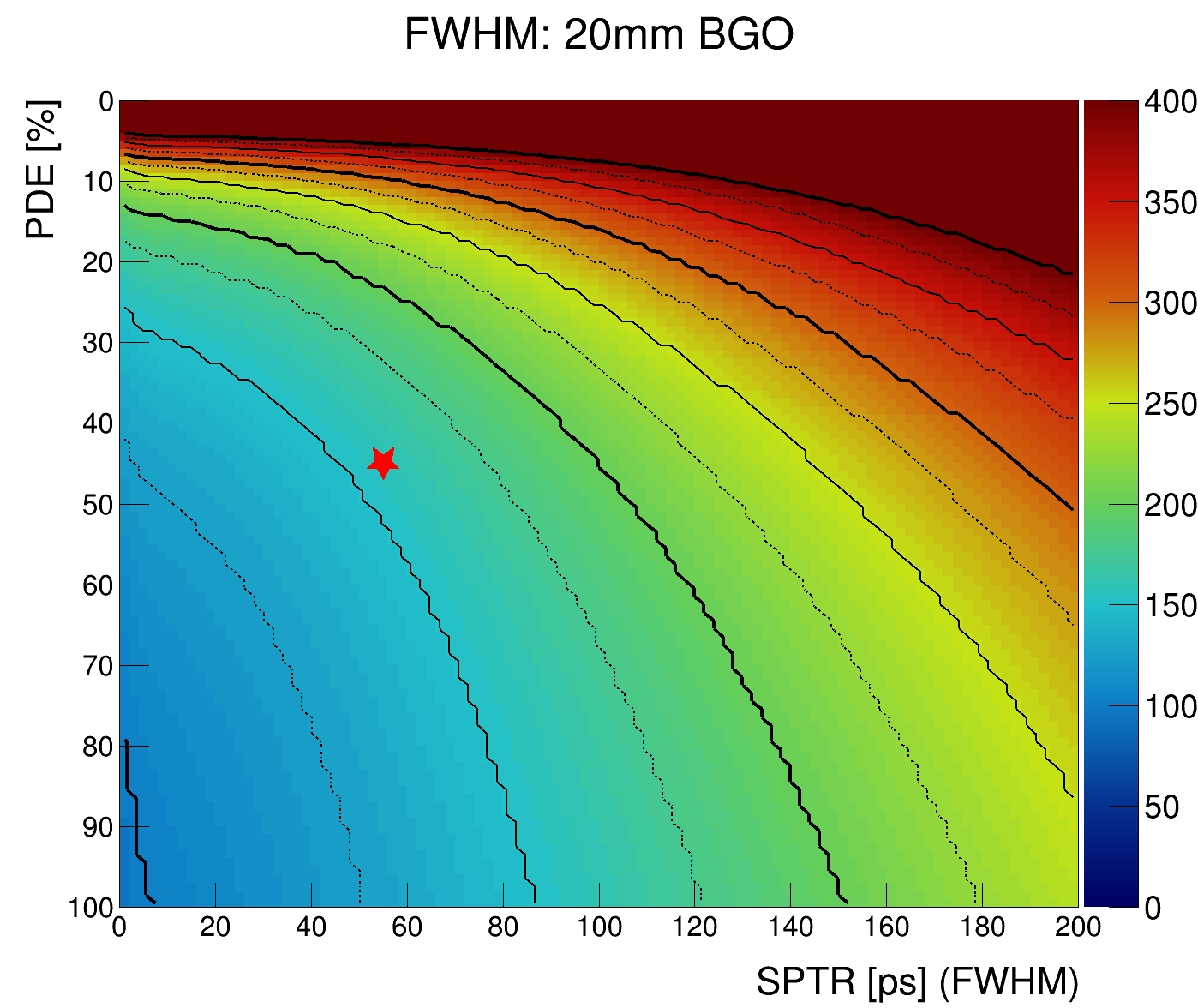} 
        \includegraphics[width=0.31\textwidth]{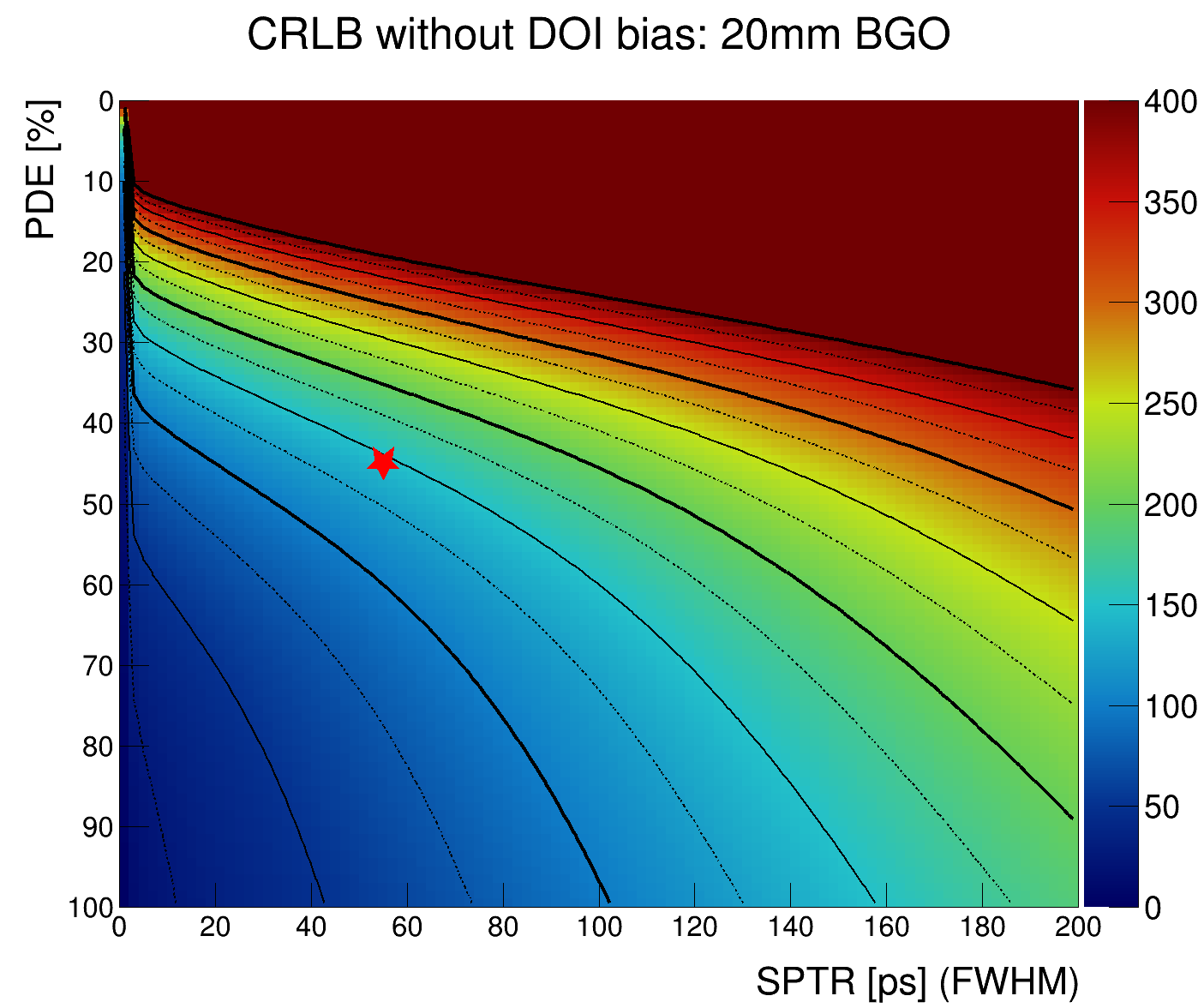} \\

         \includegraphics[width=0.31\textwidth]{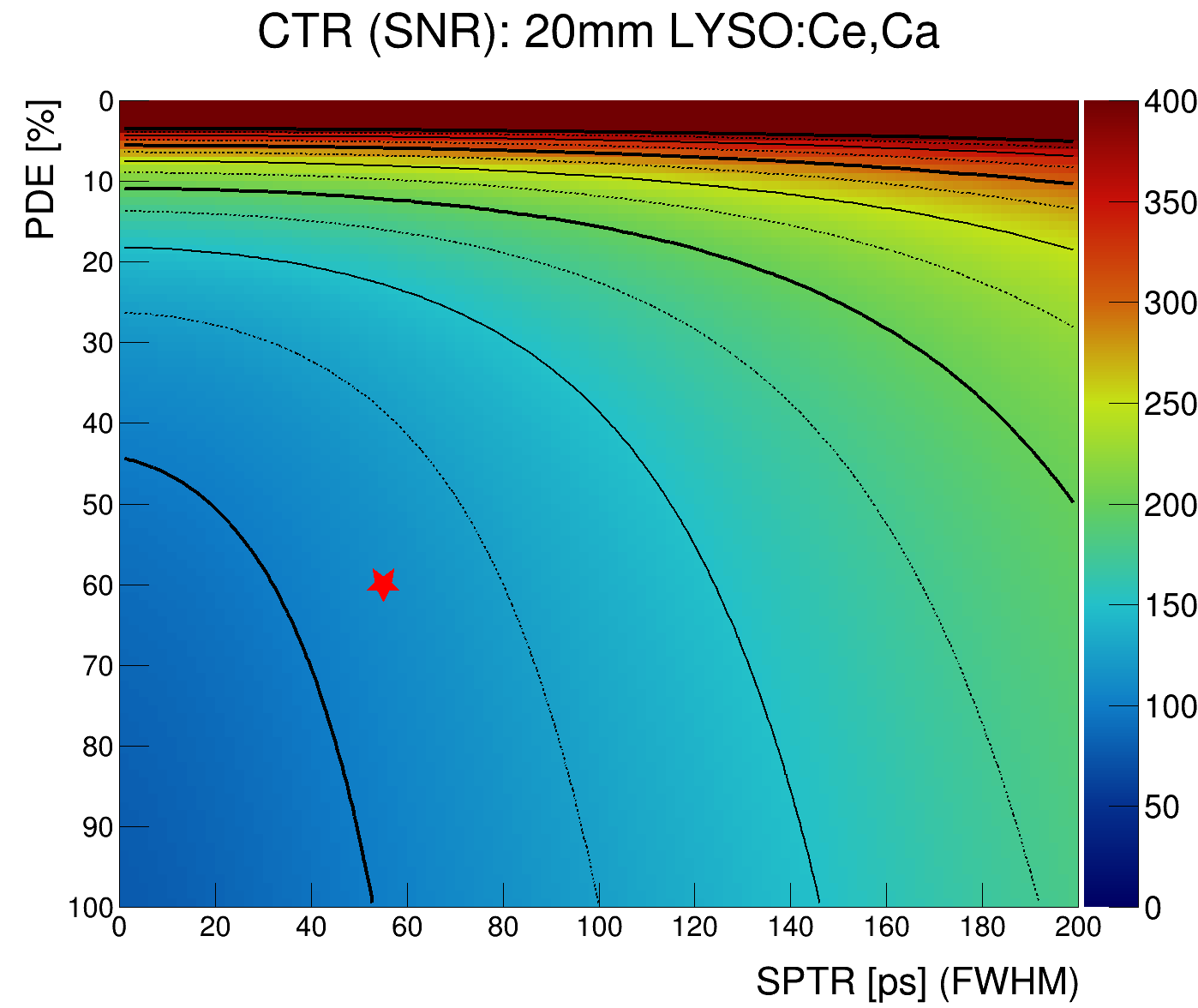}
        \includegraphics[width=0.31\textwidth]{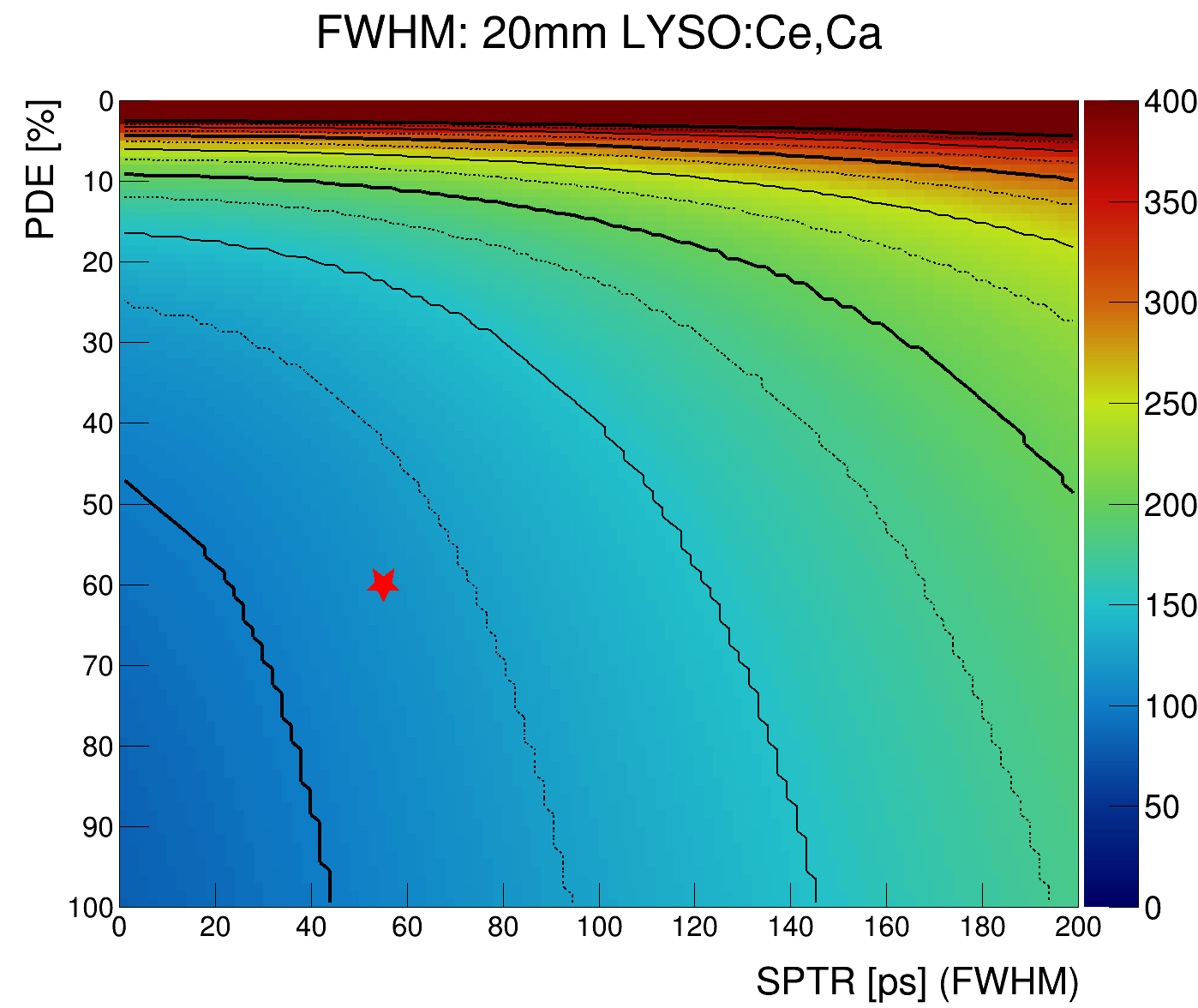} 
        \includegraphics[width=0.31\textwidth]{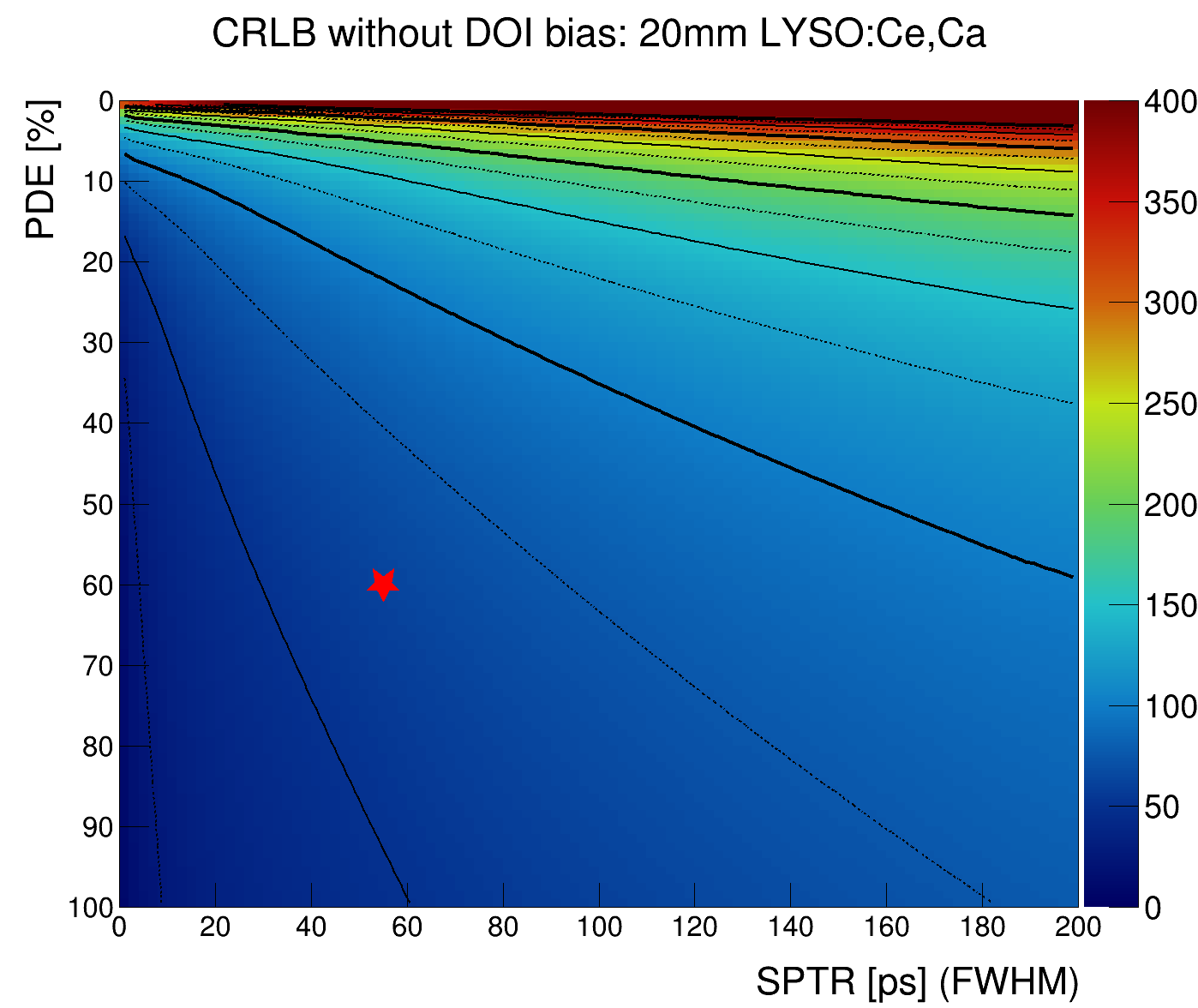} 
\end{center}
\caption{Timing metrics as function of the PDE and SPTR for 3~mm BGO (top), 20~mm BGO (center), and 20~mm thick LYSO:Ce,Ca (bottom).
Iso-lines are drawn in steps of 100~ps (thick), 50~ps (thin) and 25~ps (dotted) up to 400~ps.
The red star represents an approximation of the default configuration.}
\label{fig:SiPM-scan}
\end{figure*}
%
%With remarkable improvements of photodetectors over the last decades~\cite{Roncali-2011-ABE,ACERBI_NIMA_2019,Gola_2019_Sensors,Ota_2021_RPT,Gundacker_Heering_2020_PMB} the question arises of what is actually needed for ultra-fast timing with prompt photons.
%
%This is in particular relevant when comparing two different photodetector technologies based on single photon avalanche diodes in arrays (SiPMs), or multi channel plate photomultiplier tubes (MCP-PMTs).
%
%Among other differences, SiPMs typically have excellent photo detection efficiency of over 60\% and SPTR values of 50-100~ps FWHM have been reported~\cite{Nemallapudi_2016_JINST,Cates_2018_PMB,Merzi_2023_JINST,Gundacker_2023_PMB,Lee_2024_TRPMS}.
%
%On the other hand, MCP-PMTs may have transit time spreads (TTS, similar to SPTR) in the order of 20~ps, but generally worse detection efficiency~\cite{LEHMANN_2011_NIMA,Ota_2019_PMB,Ota_2021}.

With recent advances in photodetectors~\cite{Roncali-2011-ABE,ACERBI_NIMA_2019,Gola_2019_Sensors,Ota_2021_RPT,Gundacker_Heering_2020_PMB}, the question arises what parameters are most important for ultra-fast timing with prompt photons.
This is especially relevant when comparing SiPMs MCP-PMTs.
SiPMs offer higher detection efficiency and SPTR values of 50-100~ps FWHM~\cite{Nemallapudi_2016_JINST,Cates_2018_PMB,Merzi_2023_JINST,Gundacker_2023_PMB,Lee_2024_TRPMS}, whereas MCP-PMTs can reach transit times spread values around 20~ps but with generally lower detection efficiency~\cite{LEHMANN_2011_NIMA,Ota_2019_PMB,Ota_2021}.

We calculate the timing performance for BGO (3~mm and 20~mm) and LYSO:Ce,Ca (20~mm) as a function of the PDE (set to be equal for scintillation and Cherenkov light) and SPTR to provide an overview of photodetector impact on different fast-timing approaches.
Figure~\ref{fig:SiPM-scan} shows the calculated timing performance in terms of CTR (SNR), FWHM and CRLB with DOI information (no DOI bias).
It can be seen that the timing resolution is very metric-dependent.
For instance the FWHM for 3~mm thick BGO is mostly impacted by the SPTR and PDE changes have little influence.
The complete opposite is the case for the CTR (SNR) for 20~mm thick BGO, the SPTR only barely impacts the timing capability while PDE improvements are much more important.
The CTR (SNR) and FWHM shape is very similar for LYSO:Ce,Ca, indicating the TOF distribution to be close to a Gaussian distribution.
CTR (CRLB) values below 75~ps for 20~mm BGO and below 50~ps for 20~mm LYSO:Ce,Ca are theoretically possible if the DOI can be corrected and the photodetector exceeds $\approx60\%$ PDE with $\approx$ 30~ps SPTR.
\subsection{Limitations and future work}
A major advantage of this analytic approach over simulations is the computation time.
Despite numeric convolutions and integration steps of around $0.3~$ps, the timing metrics of a single configuration can be calculated in less than one second on a standard laptop.
For instance, the scans of 100 x 100 configurations in figure~\ref{fig:SiPM-scan} took about 1 hour each ($\approx$~0.3 seconds per configuration) and are even faster for a more aggressive truncation with LYSO mentioned in section~\ref{sec:scintillation}.
On contrary, Monte-Carlo simulations need to trace the optical path of photons for thousands of events to obtain an accurate time delay distribution or probability density function, which makes such parameter surveys very resource intensive.

This speed advantage comes with a tradeoff in accuracy and flexibility, since our model is invariant to the crystal pitch and light attenuation or surface effects~\cite{Trigila_TRPMS_2024} are not considered.
This means the information value of our model is very limited for depolished crystals, crystals with a very small pitch, or light emission close to the absorption edge.
Incorporating a compatible model for light attenuation and specular reflectors on the lateral sides of the crystal will be subject for future work to improve our framework.
An analytic re-distribution of optical photons (eg. with a diffuse reflector~\cite{Gonzalez-Montoro_2021_BPEE}) is most likely not feasible and optical simulations are the best solution to generate the PDF for those configurations.

Another limitation is that we solely consider photoelectric events and all photons are generated in one point.
Firstly, hot recoil-electrons can travel for hundreds of micrometers~\cite{Loignon-Houle_2022_NIMA} and optical photons are produced along its (stochastic) track.
Secondly, following a photoelectric interaction in the crystal with the outer electron shell, X-rays with this binding energy are released which can travel for several millimeters or may be lost.
Lastly, gamma-rays may first deposit part of its energy via Compton interaction and deposit the remaining energy at a different position via photoelectric interaction.
For all these cases it is next to impossible to generate an analytic model and simulations are the most practical approach if this level of accuracy is required.

Comparing our results with published experiments, we observe that there is very good agreement for scintillation-based detectors~\cite{Gundacker_2020_PMB}, while our model predicts much better timing values for BGO and materials with a strong Cherenkov photon signature~\cite{Kratochwil_IEEE_2021,Gundacker_2023_PMB,Piller_2024_PMB,Lee_2024_TRPMS,Kratochwil_TRPMS_2024_TlCl}.
This is attributed to the information loss due to the transition from digital-like photodetectors (this study), where each timestamp is individually extracted, to analog photodetectors, where the analog signal consists of the sum over multiple timestamps on top of correlated photodetector noise~\cite{Lee-NIMA-2024,Kratochwil_SensorsActivations_2025,Herweg_FP_2025}, electronic noise and bandwidth limitations of the signal slew rate~\cite{Gundacker_2023_PMB}.
To reduce this mismatch we foresee to extend our approach with analog photodetectors and readout electronics similar to~\cite{Gundacker_2013_JINST}.

The Fisher information and CRLB formalism is a very powerful tool to estimate the timing capabilities of a configuration (or extract the information content of other parameters such as the DOI~\cite{Loignon-Houle_2021_PMB} or Cherenkov photon number~\cite{Kratochwil_2020_PMB,Loignon-Houle_2023_TRPMS}).
In TOF-PET imaging though, other metrics like the FWHM or CTR (SNR) might be more relevant for image reconstruction.
Thus any theoretical improvement need to be carefully guided with practical timing estimators which yield actual TOF distributions to ensure the improvement is also reflected in terms of other timing metrics.
As of now, the non-Gaussian TOF distributions are mostly relevant for prompt photons.
However, we also observe mild deviations for standard scintillators, indicated for instance in table~\ref{tab:crystals} with differences between FWHM and CTR (SNR) of 2-3~ps.
These differences are more pronounced with better SPTR values, thicker crystals, fast decay times, and in general any configuration where the central limit theorem is challenged.

%\begin{comment}
%Limited attention~\cite{Efthimiou_2020_EJNMMI,Efthimiou_2021_TRPMS,Nuyts-TMI-2023} to complex TOF-shapes has been given by the TOF-PET image reconstruction community so far.
%
%Different imaging tasks may prioritize divergent timing metrics, justifying R\&D in different directions.
%
%If the noise is under control with sufficient counts, we can hypothesize that narrow FWHM is more important for direct positron emission imaging (dPEI)~\cite{Kwon_2021_NP} and tails or side peaks have little contribution~\cite{Ota_2019_PMB}.

Finally, we would like to comment on the shape of TOF distributions and the most appropriate metrics to describe them.
Different imaging tasks may prioritize divergent timing metrics, justifying R\&D in different directions.
Narrowing the FWHM may be the most important timing metric for high spatial resolution, when localization of small lesions is required, or for direct positron emission imaging~\cite{Kwon_2021_NP}.
For classic (multi kernel~\cite{Efthimiou_2020_EJNMMI,Nuyts-TMI-2023}) TOF-PET image reconstruction, limited counts, or detecting larger lesions, the reduction of tails might be a higher priority, thus more weight need to be given to reduce the CTR (SNR).
While this becomes speculative without imaging-task-specific investigation, it underlines the importance of accurately modeling the full (complex) TOF-distribution.
%\end{comment}
%
%
\section{Conclusion}
An analytic framework to calculate the timing performance of light-based radiation detectors with prompt photons and non-negligible light transport was developed.
This is an extension to existing literature and capable to generate complex TOF-distribution  which were characterized with three different timing metrics and the Cramér Rao Lower Bound formalism.
Our model reproduces the overall timing behavior of more simple approximations without light transport and prompt photons as well as Monte-Carlo simulations and is able to analytically predict more complex detector configurations.
The detection of prompt photons significantly alter the TOF-distribution characterized by narrow peaks with pronounced tails.
This model enables cost-effective surveys of factors contributing to timing capabilities , making it a powerful tool to guide future detector R\&D for fast timing applications.
%
%
%\vspace{-3mm}
\section*{Acknowledgment}
N. Kratochwil acknowledges the support of ChatGPT \text{(OpenAI)} in assisting with the understanding of mathematical concepts and the development of the underlying software (C++) for this manuscript. The authors are solely responsible for the design, analysis, and interpretation of the results. We thank Maxime Toussaint for discussions on the DOI bias. \\
The dataset used and or analyzed during the current study are available from the corresponding author on reasonable request.
All authors declare that they have no known conflicts of interest in terms of competing financial interests or personal relationships that could have an influence or are relevant to the work reported in this paper. \\
All authors gave their approval for the final version of the manuscript.
%
%\printacronyms
%
%\vspace{-3mm}
\bibliography{manuscript}

\end{document}